\documentclass[11pt]{article}

\makeatletter\renewcommand{\section}{\@startsection
{section}{1}{\z@}{-3.5ex plus -1ex minus
    -.2ex}{2.3ex plus .2ex}{\bf }}

\makeatletter\renewcommand{\subsection}{\@startsection{subsection}{2}{\z@}{-3.25ex
plus -1ex minus
   -.2ex}{1.5ex plus .2ex}{\it }}
\makeatletter\renewcommand{\subsubsection}{\@startsection{subsubsection}{3}{-2.45ex}{-3.25ex
plus -1ex minus -.2ex}{1.5ex plus .2ex}{\it }}
\renewcommand{\thesection}{\arabic{section}}
\renewcommand{\thesubsection}{\arabic{section}.\arabic{subsection}.}

\renewcommand{\theequation}{\thesection.\arabic{equation}}
\makeatletter \@addtoreset{equation}{section}

\usepackage{graphicx}
\usepackage{epsfig} 

\newcommand{\be}{\begin{equation}}
\newcommand{\ee}{\end{equation}}
\newcommand{\bea}{\begin{array}}
\newcommand{\ea}{\end{array}}
\newcommand{\beqa}{\begin{eqnarray}}
\newcommand{\eeqa}{\end{eqnarray}}
\newcommand{\nn}{\nonumber}
\newcommand{\elll}{\ell_L}
\newcommand{\ellr}{\ell_R}
\renewenvironment{thebibliography}[1]
     {\baselineskip=16pt plus 2pt minus 1pt
      \section*{\large\refname
        \@mkboth{\MakeUppercase\refname}{\MakeUppercase\refname}}%
     \list{\@biblabel{\@arabic\c@enumiv}}%
           {\settowidth\labelwidth{\@biblabel{#1}}%
            \leftmargin\labelwidth
            \advance\leftmargin\labelsep
            \@openbib@code
            \usecounter{enumiv}%
            \let\p@enumiv\@empty
            \renewcommand\theenumiv{\@arabic\c@enumiv}}%
      \sloppy
      \clubpenalty4000
      \@clubpenalty \clubpenalty
      \widowpenalty4000%
      \sfcode`\.\@m}

\let\fn\footnote
\renewcommand{\footnote}[1]{\linespread{1.1}\fn{#1}\linespread{1.29}}

\usepackage{amsfonts}
\usepackage{amssymb}
\usepackage{mathrsfs}
\usepackage{amsmath,amssymb}
\usepackage{bbm}
\usepackage{bm}

\newcommand{\appendices}{\section*{Appendix}\setcounter{subsection}{0} \setcounter{equation}{0}
\renewcommand{\thesubsection}{\Alph{subsection}.}
\renewcommand{\theequation}{\thesubsection\arabic{equation}}
}

\def\tyng(#1){\hbox{\tiny$\yng(#1)$}}

\begin{document}

\begin{titlepage}
\begin{flushright}
\end{flushright}
\vskip 2.0cm

\begin{center}

\centerline{{\Large \bf Equivariant Reduction of $U(4)$ Gauge Theory over $S_F^2 \times S_F^2$ and}} 
\vskip 1em
\centerline{{\Large \bf and the Emergent Vortices}}

\vskip 2em

\centerline{\large \bf Se\c{c}kin~K\"{u}rk\c{c}\"{u}o\v{g}lu}

\vskip 2em

\centerline{\sl Middle East Technical University,} \centerline{\sl Department of Physics,}
\centerline{\sl Dumlupinar Boulevard, 06800, Ankara, Turkey}

\vskip 1em

{\sl  e-mail:}  \hskip 2mm {\sl kseckin@metu.edu.tr} 
\end{center}

\vskip 2cm

\begin{quote}
\begin{center}
{\bf Abstract}
\end{center}
\vskip 2em

We consider a $U(4)$ Yang-Mills theory on ${\cal M} \times S_F^2 \times S_F^2 $ where ${\cal M}$ is an arbitrary Riemannian manifold and $S_F^2 \times S_F^2$ is the product of two fuzzy spheres spontaneously generated from a $SU(\cal {N})$ Yang-Mills theory on ${\cal M}$ which is suitably coupled to six scalars in the adjoint of $U({\cal N})$. We determine the  ${\rm SU(2)} \times {\rm SU(2)} $-equivariant $U(4)$ gauge fields and perform the dimensional reduction of  the theory over $S_F^2 \times S_F^2$. The emergent model is a $U(1)^4$ gauge theory coupled to four complex and eight real scalar fields. We study this theory on ${\mathbb R}^{2}$ and find that, in certain limits, it admits vortex type solutions with $U(1)^3$ gauge symmetry and discuss some of their properties.

\vskip 5pt

\end{quote}

\end{titlepage}

\setcounter{footnote}{0}

\newpage

\section{Introduction}

Recently, there has been significant advances in understanding the structure of gauge theories possessing fuzzy extra dimensions \cite{Aschieri:2003vy, Aschieri:2006uw} (for a review on fuzzy spaces see \cite{Book}). It is known that in certain $SU({\cal N})$ Yang-Mills theories on a manifold ${\cal M}$, which are suitably coupled to a set of scalar fields, fuzzy spheres may be generated as extra dimensions by spontaneous symmetry breaking. The vacuum expectation values (VEVs) of the scalar fields form the fuzzy sphere(s), while the fluctuations around the vacuum are interpreted as gauge fields over $S_F^2$ or $S_F^2 \times S_F^2$ \cite{Aschieri:2006uw, Steinacker3}. The resulting theories can therefore be viewed as gauge theories over $M \times S_F^2$ and $ M \times S_F^2 \times S_F^2$ with smaller gauge groups; which is further corroborated by the expansion of a tower of Kaluza-Klein modes of the gauge fields. Inclusion of fermions into this theory was considered in \cite{Steinacker3, Steinacker2}. For instance, in \cite{Steinacker2} an appropriate set of fermions in $6D$ allowed for an effective description of Dirac fermions on $M^4 \times S_F^2$, which was further affirmed by a Kaluza-Klein modes expansion over $S_F^2$. It was also found that a chirality constraint on the fermions leads to a description in terms of "mirror fermions" in which each chiral fermion comes with a partner with opposite chirality and quantum numbers. 
 
It appears well motivated to investigate equivariant parametrization of gauge fields and perform dimensional reduction over the fuzzy extra dimensions to shed some further light into the structure of these theories. Essentially, it is possible to use the well known coset space dimensional reduction (CSDR) techniques to achive this task. To briefly recall the latter consider a Yang-Mills theory with a gauge group $S$ over the product space ${\cal M} \times G/H$. $G$ has a natural action on its coset, and requiring the Yang-Mills gauge fields to be invariant under the $G$ action up to $S$ gauge transformations leads to a G-equivariant parametrization of the gauge fields and subsequently to the dimensional reduction of the theory after integrating over the coset space $G/H$ \cite{Forgacs, Zoupanos}. CSDR techniques have been widely used as a method in attempts to obtain the standard model on the Minkowski space $M^4$ starting from a Yang-Mills-Dirac theory on the higher dimensional space $M^4 \times G/H$ (for a review on this topic reader can consult \cite{Zoupanos}). The widely known, prototype example of CSDR is the $SU(2)$-equivariant reduction of the Yang-Mills theory over $\mathbb{R}^4$ to an abelian Higgs model on the two-dimensional hyperbolic space $\mathbb{H}^2$, which was formulated by Witten \cite{Witten} prior to the development of the formal approach of \cite{Forgacs}, and it led to the construction of instanton solutions with charge greater than $1$. 

Another approach, parallel to the CSDR scheme, using the language of vector bundles and quivers is also known in the literature \cite{Prada}. In recent times, this approach has been employed in a wide variety of problems, including the formulation of quiver gauge theory of non-Abelian vortices over ${\mathbb R}^{2d}_\theta$ corresponding to instantons on ${\mathbb R}^{2d}_\theta \times S^2$, ${\mathbb R}^{2d}_\theta \times S^2 \times S^2$ \cite{Popov-Szabo, Lechtenfeld1}, to the construction of vortex solutions over Riemann surfaces which become integrable for appropriate choice of the parameters \cite{Popov} and to the construction of non-Abelian monopoles over  ${\mathbb R}^{1,1} \times S^2$ in \cite{Popov2}. In  \cite{Dolan-Szabo}, reduction of the Yang-Mills-Dirac theory on $M \times S^2$ is considered with a particular emphasis on the effects of the non-trivial monopole background on the physical particle spectrum of the reduced theory. Dimensional reduction over quantum sphere is recently studied and led to the formulation of q-deformed quiver gauge theories and non-Abelian q-vortices \cite{Landi-Szabo}.

Both of these techniques have also been applied to Yang-Mills theories over ${\mathbb R}^{2d}_\theta \times S^2$ \cite{Lechtenfeld:2003cq}, where ${\mathbb R}^{2d}_\theta$ is the $2d$ dimensional Groenewald-Moyal space; a prime example of a noncommutative space. In this framework, Donaldson-Uhlenbeck-Yau (DUY) equations of a $U(2k)$ Yang-Mills theory have been reduced to a set of equations on ${\mathbb R}^{2d}_\theta$ whose solutions are given by BPS vortices on ${\mathbb R}^{2d}_\theta$ and the properties of the latter have been elaborated. 

Starting with the article \cite{Seckin-Derek}, we have initiated investigations on the equivariant reduction of gauge theories over fuzzy extra dimensions. In \cite{Seckin-Derek} the most general $SU(2)$-equivariant $U(2)$ gauge field over ${\cal M} \times S_F^2$ have been found, and it was utilized to perform the dimensional reduction over $S_F^2$. It was shown that for ${\cal M} = {\mathbb R}^2$ the emergent theory is an Abelian Higgs type model which has non-BPS vortex solutions corresponding to the instantons in the original theory. There it was also found that these non-BPS vortices attract or repel depending on the parameters in the model. This article has been followed up by investigating the situation in which ${\cal M}$ is also a noncommutative space \cite{seckin-PRD}. Performing the ${\rm SU(2)}$-equivariant dimensional reduction of this theory led to a noncommutative $U(1)$ theory which couples adjointly to a set of scalar fields. On the Groenewald-Moyal plane ${\cal M} = {\mathbb R}^{2}_\theta$ the emergent models admit noncommutative vortex as well as fluxon solutions, which are non-BPS and devoid of a smooth commutative limit as $\theta \rightarrow 0$.

As we have noted earlier, gauge theory on $M^4 \times S_F^2 \times S_F^2$ has been recently investigated in \cite{Steinacker3}. For this purpose authors of \cite{Steinacker3} have considered a $SU({\cal N})$ gauge theory on $M^4$, which is suitably coupled to six scalar fields in the adjoint of $U({\cal N})$. The model has the same field content as that of the bosonic part of the $N=4$ SUSY Yang-Mills theory, but comes together with a potential breaking the $N=4$ supersymmetry and the $R$-symmetry which is a global $SU(4)$. The deformed potential makes possible (after spontaneous symmetry breaking) the identification of the VEV's of the scalars with $S_F^2 \times S_F^2$ and the fluctuations around this vacuum as gauge fields on $S_F^2 \times S_F^2$. Structure of fermions in this theory is elaborated in \cite{Steinacker3}. In a related article, it was shown that twisted fuzzy spheres can be dynamically generated as extra dimensions starting from a certain orbifold projection of a $N = 4$ SYM theory whose consequences have been discussed in \cite{Zoupanos-1}. For a review on these results \cite{Zoupanos-Review} can be consulted. 

In the present article, we investigate the ${\rm SU(2)} \times {\rm SU(2)} $ equivariant formulation of a $U(4)$ gauge theory over $S_F^2 \times S_F^2$. Starting from the $SU({\cal N})$ gauge theory model described above, but now put on some Riemannian Manifold ${\cal M}$, we focus on a $U(4)$ gauge theory on ${\cal M} \times S_F^2 \times S_F^2 $  after spontaneous symmetry breaking. We determine the  ${\rm SU(2)} \times {\rm SU(2)} $-equivariant $U(4)$ gauge fields and perform the dimensional reduction of  the theory over $S_F^2 \times S_F^2$. The emergent model is a $U(1)^4$ gauge theory coupled to four complex and eight real scalar fields. We study this theory on ${\mathbb R}^{2}$ and find that, in certain limits, it admits vortex type solutions with $U(1)^3$ gauge symmetry and discuss some of their properties.

Our work in the rest of the paper is organized as follows. In section 2, we give the basics of the $SU(\cal{N})$ gauge theory over ${\cal M}$ and indicate how the gauge theory over ${\cal M}$ dynamically develops $S_F^2 \times S_F^2$ as extra dimensions. This is followed by a systematic construction of the ${\rm SU(2)} \times {\rm SU(2)} $-equivariant $U(4)$ gauge field using essentially the $SO(4) \approx {\rm SU(2)} \times {\rm SU(2)}$ representation theory. In section 3, we present the results of the equivariant reduction over ${\cal M} \times S_F^2 \times S_F^2$ and give the reduced action in full, and find that the emergent model is a $U(1)^4$ gauge theory coupled to four complex and eight real scalar fields. This is ensued by a discussion of the structure of the reduced action. In section 4, we present non-trivial solutions of the reduced action on ${\mathbb R}^{2}$ for two different limiting cases of the parameters $a_L$ and $a_R$ in the theory and demonstrate that, these particular models have vortex solutions with $U(1)^3$ gauge symmetry which tend to attract or repel  at the critical point of the parameter space $g {\tilde g} = 1$. For completeness, brief definitions of $S_F^2$ and
$S_F^2 \times S_F^2$ are given in appendix A and basics of the $U(\cal{N})$ gauge theory over ${\cal M} \times S_F^2$ and the $U(2)$-equivariant gauge field parametrization are discussed in appendix B. In appendix C, we collect the explicit expressions after dimensional which is presented in section 3.

\section{$U(4)$ Gauge Theory over ${\cal M} \times S_F^2 \times S_F^2$}

{\it  i. Gauge theory on ${\cal M} \times S_F^2 \times S_F^2$:}

\vskip 1em

We start with an $SU({\cal N})$ gauge theory coupled adjointly to six scalar fields $\Phi_i \,, (i = 1\,, \cdots\,, 6)$. The relevant action is given in the form \cite{Steinacker3}
\be
S = \int_{{\cal M} } \, \mbox{Tr}_{{\cal N}} \Big (\frac{1}{4g^2} F_{\mu \nu}^\dagger F_{\mu \nu} + (D_\mu \Phi_i)^\dagger (D_\mu \Phi_i) \Big ) + V(\Phi) \,.
\label{eq:actionsun}
\ee
In this expression, $A_\mu$ are $su({\cal N})$ valued anti-Hermitian gauge fields, $\Phi_i \,(i=1, \cdots 6)$ are six anti-Hermitian scalars transforming in the adjoint of ${\rm SU(}{\cal N}{\rm )}$ and $D_\mu \Phi_i  = \partial_\mu \Phi_i + \lbrack A_\mu \,, \Phi_i\rbrack$ are the covariant derivatives.

It is assumed further that $\Phi_i \,, (i = 1 \,,\cdots\,, 6)$ transform in the vector representation of a global 
$SU(4) \cong SO(6)$ group.

When considered on the four dimensional Minkowski spacetime $M^4$, depending on the form of the potential term $V(\Phi)$, the action (\ref{eq:actionsun}) corresponds to the bosonic part of the $N = 4$ super Yang-Mills theory with the global $SU(4)$ being its $R$-symmetry, or a modification of it thereof. The potential may have the form 
\be
V(\Phi) = V_{N= 4}(\Phi) + V_{break}(\Phi) \,,
\ee
where the first term corresponds to the potential of the $N = 4$ super Yang-Mills theory
\be
V_{N = 4}(\Phi) = \frac{1}{4} g_4^2 \sum_{i\,,j}^6 \lbrack \Phi_i \,, \Phi_j \rbrack^2 \,,
\ee
while the second term breaks both the $N = 4$ supersymmetry and the $R$-symmetry. It also worths to mention that the above action (\ref{eq:actionsun}) descends from a ten-dimensional $N = 1$ super Yang-Mills theory by dimensional reduction. We will not  review this here as it is not necessary for our purposes, however a quick discussion can be found in \cite{Steinacker3}.

We would like to see now how the product of two fuzzy spheres emerges as extra dimensions from this theory as a consequence of spontaneous breaking of the original gauge symmetry. Following the discussion in \cite{Steinacker3}, we consider a potential of the form
\be
V(\Phi) = \frac{1}{g_L^2} V_1(\Phi^L) + \frac{1}{g_R^2} V_1(\Phi^R) + \frac{1}{g_{LR}^2} V_1(\Phi^{L,R})
+ a_L^2 V^L_2(\Phi_L) + a_R^2 V^R_2(\Phi_R) \,,
\ee
where 
\be
\Phi_a^L = \Phi_a \,, \quad \Phi_a^R = \Phi_{a+3} \,,  \quad (a = 1,2,3) \,,
\ee
and 
\begin{gather}
V_1(\Phi^L) = \mbox{Tr}_{{\cal N}} F_{ab}^{L  \dagger} F_{ab}^L  \,, \quad F_{ab}^L= \lbrack \Phi_a^L \,, \Phi_b^L \rbrack - \varepsilon_{abc} \Phi_c^L \nn \\
V_1(\Phi^R) = \mbox{Tr}_{{\cal N}} F_{ab}^{R  \dagger} F_{ab}^R  \,, \quad F_{ab}^R= \lbrack \Phi_a^R \,, \Phi_b^R \rbrack - \varepsilon_{abc} \Phi_c^R \nn \\
V_2(\Phi^L) = \mbox{Tr}_{{\cal N}} (\Phi^L_a \Phi^L_a + {\tilde b}_L)^2  \,, \quad
V_2(\Phi^R) = \mbox{Tr}_{{\cal N}} (\Phi^R_a \Phi^R_a + {\tilde b}_R)^2 \nn \\
V_1(\Phi^{L,R}) = \mbox{Tr}_{{\cal N}}  F_{ab}^{(L \,, R) \dagger} F_{ab}^{(L \,, R)} \,, \quad F_{ab}^{(L \,, R)} = \lbrack \Phi_a^L \,, \Phi_b^R \rbrack  \,.
\end{gather}

We observe that the potential $V(\Phi)$ is positive definite, and it is possible to pick ${\tilde b}_L$ and ${\tilde b}_R$ as the quadratic Casimirs of respectively ${\rm SU(2)}_L$ and ${\rm SU(2)}_R$ with IRR's labeled by $\ell_L$ and $\ell_R$ 
\be
{\tilde b}_L = \ell_L (\ell_L + 1)\,, \quad {\tilde b}_R = \ell_R (\ell_R + 1)\,, \quad 2\ell_L\,, \, 2\ell_R \in \mathbb{Z} \,.
\ee
If it is further assumed that ${\cal N} = (2 \ell_L +1) (2 \ell_R +1) n$, ($n \in {\mathbb Z}$), then the configuration
\beqa
\Phi_a^L &=& X_a^{(2 \ell_L + 1)} \otimes {\bf 1}_{(2 \ell_R + 1)}  \otimes {\bf 1}_n \,, \nn \\
\Phi_a^R &=& {\bf 1}_{(2 \ell_L + 1)} \otimes X_a^{(2 \ell_R + 1)}  \otimes {\bf 1}_n \,,
\label{eq:minima}
\eeqa
\be
\lbrack \Phi_a^L \,, \Phi_b^R \rbrack = 0  \,,
\ee
is a global minimum of the potential $V(\Phi)$ where $X_a^{(2 \ell_L + 1)}$ and $X_a^{(2 \ell_R + 1)}$ are the anti-Hermitian generators of ${\rm SU(2)}_L$ and ${\rm SU(2)}_R$ respectively in the IRR's $\ell_L$ and $\ell_R$, with the commutation relations
\be
\lbrack  X_a^{(2 \ell_L + 1)} \,, X_b^{(2 \ell_L + 1)} \rbrack =  \varepsilon_{abc} X_c^{(2 \ell_L + 1)} \,, \quad \lbrack X_a^{(2 \ell_R + 1)} \,, 
X_b^{(2 \ell_R + 1)} \rbrack =  \varepsilon_{abc} X_c^{(2 \ell_R + 1)} \,.
\ee
This vacuum configuration spontaneously breaks the ${\rm SU}({\cal N})$ down to ${\rm U}(n)$ which is the commutant of $\Phi_a^L\,, \Phi_a^R$ in (\ref{eq:minima}). 

Defining
\begin{gather}
{\hat x}_a^L = \frac{i}{\sqrt{\ell_L (\ell_L + 1)}} X_a^{(2 \ell_L + 1)} \otimes 1_{(2 \ell_R +1)} \,, \quad 
{\hat x}_a^R = 1_{(2 \ell_L +1)} \otimes \frac{i}{\sqrt{\ell_R (\ell_R + 1)}} X_a^{(2 \ell_R + 1)} \,, \\
{\hat x}_a^L {\hat x}_a^L = 1 \,, \quad {\hat x}_a^R {\hat x}_a^R = 1 \,.
\end{gather}
the vacuum is a product of two fuzzy spheres $S^2_F \times S_F^2$ generated by ${\hat x}_a^L $ and ${\hat x}_a^R$. (see appendix A for a description of $S^2_F \times S_F^2$).

Fluctuations about this vacuum give a $U(n)$ gauge theory over $S_F^2 \times S_F^2$. We can write
\be
\Phi_a^L = X_a^L + A_a^L \,, \quad \Phi_a^R = X_a^R + A_a^R 
\label{eq:config1old} 
\ee
where $A_a^L \,, A_a^R \in u(2\ell_L+1) \otimes u(2\ell_R+1) \otimes  u(n)$ with the short-hand notation 
$X_a^{(2 \ell_L+ 1)} \otimes {\bf 1}_{(2 \ell_R + 1)} \otimes {\bf 1}_n =: X_a^L$ and ${\bf 1}_{(2 \ell_L + 1)} \otimes X_a^{(2 \ell_R + 1)}  \otimes {\bf 1}_n =: X_a^R$.

Thus, $\Phi_a^L \,, \Phi_a^R $ are the ``covariant coordinates'' on $S_F^2 \times S_F^2$, and the associated
curvatures $F_{ab}^L$, $F_{ab}^R$, $F_{ab}^{L\,, R}$ take their familiar form after expanding according to (\ref{eq:config1old})
\beqa
F_{ab}^L &=& \lbrack X_a^L \,, A_b^L \rbrack - \lbrack X_b^L \,, A_a^L \rbrack + \lbrack A_a^L \,, A_b^L \rbrack - \varepsilon_{abc} A_c^L \,, \nn \\
F_{ab}^R &=& \lbrack X_a^R \,, A_b^R \rbrack - \lbrack X_b^R \,, A_a^R \rbrack + \lbrack A_a^R \,, A_b^R \rbrack - \varepsilon_{abc} A_c^R \,, \nn \\
F_{ab}^{L \,, R} &=& \lbrack X_a^L \,, A_b^R \rbrack - \lbrack X_b^R \,, A_a^L \rbrack + \lbrack A_a^L \,, A_b^R \rbrack \,.
\eeqa
Therefore, we can interpret the spontaneously broken theory as a $U(n)$ gauge theory on ${\cal M} \times S_F^2 \times S_F^2$ with $A_M := ( A_\mu \,, A_a^L \,, A_a^R)$ as the gauge fields and $F_{MN}$ as the corresponding field strength. The $V_2^L$ and the $V_2^R$ serve as constraint terms to suppress the normal components of the gauge fields on each of the fuzzy spheres, in similar manner as discussed for the case of a single fuzzy sphere in \cite{Aschieri:2006uw, Seckin-Derek}.

It is important to point out that, this gauge theory can be called the ``standard" Yang-Mills theory on ${\cal M} \times S_F^2 \times S_F^2$
if we take $g_L = g_R = {\sqrt{2}} g_{L,R} := {\tilde g}$, scale the scalar fields as ${\tilde \Phi}= {\sqrt{2}} {\tilde g} \Phi_i$ and take  
${\tilde g} g =1$, since only then it takes the form of the $L^2$ norm of $F_{MN}$. 

We also note for future use that, with the developments above
\be
\mbox{Tr}_{{\cal N}} = \frac{1}{n (2 \ell_L +1) (2 \ell_R +1)} \mbox {Tr}_{\mbox{Mat} (2 \ell_L +1)} \otimes  \mbox {Tr}_{\mbox{Mat} (2 \ell_R +1)} \otimes \mbox {Tr}_{\mbox{Mat}(n)}
\ee
where $\mbox{Mat}(k)$ denotes the algebra of $k \times k$ matrices.

Finally, it is also useful to remark that there are other possibilities for the vacuum configuration as discussed in \cite{Steinacker3} which for instance lead to $S_F^2 \times S_F^2$ carrying magnetic fluxes under the $U(1)$ component of the unbroken gauge group $SU(n) \times SU(m) \times U(1)$ after spontaneous symmetry breaking.

\vskip 1em

{\it ii. The ${\rm SU(2) \times SU(2)}$-Equivariant Gauge Field}

\vskip 1em

We will now formulate the $SU(2)_L \times SU(2)_R \cong SO(4)$-equivariant, $U(4)$ gauge theory on ${\cal M} \times
S_F^2 \times S_F^2$. The gauge fields carry the fundamental representation of $U(4)$.  We introduce $SO(4)$ symmetry generators under which $A_\mu$ is a scalar up to a $U(4)$ gauge transformation, that is carrying the $SO(4)$ IRR $(0,0)$ and $A_a^L$ and $A_a^R$ are $SO(4)$ tensors carrying the IRRs $(1,0)$ and $(0,1)$, respectively. In other words, $A_a^L$ is a vector under the $SU(2)_L$ and a scalar under the $SU(2)_R$, whereas $A_a^R$ is an $SU(2)_R$ vector and an $SU(2)_L$ scalar.

On $S_F^2 \times S_F^2$ the $SU(2) \times SU(2) \cong SO(4)$ rotational symmetry is implemented by the adjoint actions
$\mbox{ad} X_a^L$ and $\mbox{ad} X_a^R$ (see appendix A):
\be
\mbox{ad} X_a^L \cdot  = \lbrack X_a^L \,, \cdot \rbrack \,, \quad  \mbox{ad} X_a^R \cdot  = \lbrack X_a^R \,, \cdot \rbrack \,, \quad
\lbrack \mbox {ad} X_a^L \,, \mbox{ad} X_a^R \rbrack = 0 \,.
\ee
 
Let's introduce the anti-Hermitian symmetry generators
\beqa
\omega_a^L &=&  X_a^{(2 \ell_L + 1)} \otimes {\bf 1}_{(2 \ell_R + 1)}  \otimes {\bf 1}_4 - 1_{(2 \ell_L + 1)} \otimes {\bf 1}_{(2 \ell_R + 1)}  \otimes 
i \frac{L_a^L}{2}  \,, \nn \\
\omega_a^R &=&  {\bf 1}_{(2 \ell_L + 1)} \otimes X_a^{(2 \ell_R + 1)}  \otimes {\bf 1}_4 - 1_{(2 \ell_L + 1)} \otimes {\bf 1}_{(2 \ell_R + 1)}  \otimes i \frac{L_a^R}{2}  \,. 
\label{eq:SO4symm}
\eeqa
Here $L_a^L$ and $L_a^R$ are $4 \times 4$ matrices whose structure will be given shortly. They are chosen so that $\omega_a^L$ and 
$\omega_a^R$ fulfill the consistency conditions
\beqa
\lbrack \omega_a^L \,, \omega_b^L \rbrack &=& i \varepsilon_{abc} \omega_c^L \,, \nn \\
\lbrack \omega_a^R \,, \omega_b^R \rbrack &=& i \varepsilon_{abc} \omega_c^R \,, \\
\lbrack \omega_a^L \,, \omega_b^R \rbrack &=& 0 \,.
\eeqa

In order to write down the matrices $L_a^L$ and $L_a^R$ consider first the $4 \times 4 $ matrices denoted as $e_{mn} (m,n=1,2,3,4)$ whose all entries are zero except the entry on the $m^{th}$ row and the $n^{th}$ column which is $1$. We let
\be
J_a = - i \varepsilon_{abc} e_{bc} \,, \quad K_a = -i (e_{a4} - e_{4a}) \,,
\ee
and define
\be
L_a^L = J_a + K_a \,, \quad L_a^R = J_a - K_a \,.
\ee
These matrices fulfill
\beqa
\lbrack L_a^L \,, L_b^L \rbrack &=&  2 i \varepsilon_{abc} L_c^L \,, \nn \\
\lbrack L_a^R \,, L_b^R \rbrack &=&  2 i \varepsilon_{abc} L_c^R \,, \nn \\
\lbrack L_a^L \,, L_b^R \rbrack &=& 0 \,.
\eeqa
Therefore we have altogether six anti-symmetric $SU(4)$ matrices generating the two subgroups $SU(2)_L$ and $SU(2)_R$.
Remaining nine symmetric generators of $SU(4)$ may be taken as $L_a^L L_b^R$. Together with the $4 \times 4$ identity ${\bm 1}_4$,
$L_a^L$, $L_a^R$ and $L_a^L L_b^R$ span $U(4)$ and furnish a basis for the fundamental representation of $U(4)$. 

$L_a^L$ and $L_a^R$ form a $4 \times 4$ basis of the Lie algebra $so(4) = su(2) \oplus su(2)$. In addition, these matrices satisfy the relations
\beqa
L_a^L L_b^L &=& i \varepsilon_{abc} L_c^L + \delta_{ab} {\bm 1}_4 \,, \nn \\
L_a^R L_b^R &=& i \varepsilon_{abc} L_c^R + \delta_{ab} {\bm 1}_4 \,,
\eeqa
which permits to view them as two sets of $4 \times 4$`Pauli Matrices".

From the point of view of the $SO(4)$ representation theory $L_a^L$, $L_a^R$ carry the reducible representations of $SU(2)_L$ and 
$SU(2)_R$. $L_a^L$ carries two copies of the IRR $(\frac{1}{2} \,, 0)$, whereas $L_a^R$ carries two copies of the IRR $(0 \,, \frac{1}{2})$, which can be clearly observed from their Casimir operators with the eigenvalues $3$. 

As the gauge fields $A_a^L$ and $A_a^R$ on $S_F^2 \times S_F^2$ are $u(4)$ valued, they are elements of $u(2 \ell_L +1) \times u(2 \ell_R +1) \times u(4)$. Therefore, it is now clear that $L_a^L$ and $L_a^R$ in (\ref{eq:SO4symm}) are responsible for generating the $U(4)$ gauge symmetry in $SO(4)$. 

The $SU(2) \times SU(2) \cong SO(4)$-equivariance conditions stated at the beginning of this section can now be explicitly described
as the fulfillment of the following conditions under the adjoint actions of $\omega^L$ and $\omega^R$.
\beqa
&& \lbrack \omega_a^L \,, A_\mu \rbrack = 0 = \lbrack \omega_a^R \,, A_\mu \rbrack \,, \nn \\
&& \lbrack \omega_a^L \,, A_b^L \rbrack =  \varepsilon_{abc} A_c^L \,, \nn \\
&& \lbrack \omega_a^R \,, A_b^R \rbrack =  \varepsilon_{abc} A_c^R \,, \nn \\
&& \lbrack \omega_a^L \,, A_b^R \rbrack = 0 = \lbrack \omega_a^R \,, A_b^L \rbrack \,.
\eeqa
It is necessary to find explicit parametrizations of $A_\mu$, $A_a^L$ and $A_a^R$ fulfilling these conditions. The adjoint actions
of $\omega^L$ and $\omega^R$ expand in Clebsch-Gordan series as
\beqa
2 \times \left \lbrack (\ell_L\,, 0) \otimes (\frac{1}{2} \,, 0) \right \rbrack \otimes \left \lbrack (\ell_L \,, 0) \otimes (\frac{1}{2} \,, 0) \right \rbrack 
&=& 4 (0 \,, 0) \oplus 8 (1 \,, 0) \oplus \cdots \,, \\
2 \times \left \lbrack (0 \,, \ell_R) \otimes (0 \,, \frac{1}{2}) \right \rbrack \otimes \left \lbrack (0 \,, \ell_R) \otimes (0 \,, \frac{1}{2}) \right \rbrack 
&=& 4 (0 \,, 0) \oplus 8 (0 \,, 1) \oplus \cdots \,,
\eeqa
where the factor of two in each line above is due to the two copies of the IRRs $(\frac{1}{2} \,, 0)$ and $(0 \,, \frac{1}{2})$
in $L_a^L$ and $L_a^R$, respectively. Therefore the relavant part of the Clebsch-Gordan expansion takes the form
\be
4 (0 \,, 0) \oplus 8 (1 \,, 0) \oplus  8 (0 \,, 1) \,.
\ee
The solution space for $A_\mu$ is then $4$-dimensional, whereas each of the solution spaces of $A_a^L$ and $A_a^R$ are $8$-dimensional.

It is not very hard to see that there are four invariants under the action of $\omega_a^L$ and $\omega_a^R$. These are the three `idempotents"
\beqa
Q_L  &=& \frac{X_a^{\ell_L} \otimes {\bm 1}_{(2\ell_R +1)} \otimes L_a^L - \frac{i}{2} {\bf 1}}{\ell_L+1/2}  \,, \quad Q_L^\dagger = - Q_L \,, \quad Q_L^2 = -{\bm 1}_{4(2\ell_L+1)(2 \ell_R +1)} \,, \\
Q_R &=& \frac{{\bm 1}_{(2\ell_L +1)}  \otimes  X_a^{\ell_R} \otimes L_a^R - \frac{i}{2} {\bf 1}}{\ell_R+1/2}  \,, \quad Q_R^\dagger = - Q_R \,, \quad Q_R^2 = -{\bm 1}_{4(2\ell_L+1)(2 \ell_R +1)} \,, 
\eeqa
\begin{multline}
i Q_L Q_R = i \frac{(X_a^{\ell_L} \otimes {\bm 1}_{(2\ell_R +1)} \otimes L_a^L - \frac{i}{2} {\bf 1})({\bm 1}_{(2\ell_L +1)}  \otimes X_a^{\ell_R} \otimes L_a^R - \frac{i}{2} {\bf 1})}{(\ell_L+1/2)({\ell_R+1/2})} \,, \\[0.8 em] 
( i Q_L Q_R)^\dagger = - i Q_L Q_R \,, \quad   (i Q_L Q_R)^2 = - {\bm 1}_{4(2\ell_L+1)(2 \ell_R +1)} \,,
\end{multline}
which are all $\lbrack 4(2\ell_L+1)(2 \ell_R +1) \rbrack^2$ matrices and the identity matrix $-{\bm 1}_{4(2\ell_L+1)(2 \ell_R +1)}$.

These lead to the parametrization
\be
A_\mu = \frac{1}{2} a_\mu^L Q^L + \frac{1}{2} a_\mu^R Q^R + \frac{i}{2} b_\mu {\bm 1} + \frac{1}{2} i c_\mu Q^L Q^R  \,,
\ee
where $a_\mu$, $b_\mu$, $c_\mu$ and $d_\mu$ are all Hermitian $U(1)$ gauge fields, and to the parametrizations
\begin{multline}
A_a^L = \frac{1}{2} (\chi_1 + \chi_1^\prime) [X_a^L\,, Q^L] + \frac{1}{2} (\chi_2 + \chi_2^\prime - 1) Q^L [X_a^L \,,Q^L] + i \frac{1}{2} \chi_3 \frac{1}{2} \{ \widehat{X}_a^L \,, Q^L \} + \frac{1}{2} \chi_4 \widehat{\omega}_a^L \\
+ \frac{1}{2} (\chi_1 - \chi_1^\prime) i Q^R [X_a^L\,, Q^L] + \frac{1}{2} (\chi_2 - \chi_2^\prime) i Q^R Q^L [X_a^L \,,Q^L] + i \frac{1}{2} \chi_3^\prime \frac{1}{2} i Q^R \{ \widehat{X}_a^L \,, Q^L \} + \frac{1}{2} \chi_4^\prime i Q^R \widehat{\omega}_a^L \,.
\end{multline}
\begin{multline}
A_a^R = \frac{1}{2} (\lambda_1 + \lambda_1^\prime)  [X_a^R\,, Q^R] + \frac{1}{2} (\lambda_2 + \lambda_2^\prime- 1) Q^R [X_a^R \,,Q^R] + i \frac{1}{2} \lambda_3 \frac{1}{2} \{ \widehat{X}_a^R\,, Q^R \} + \frac{1}{2} \lambda_4 \widehat{\omega}_a^R \\
+ \frac{1}{2} (\lambda_1 - \lambda_1^\prime) i Q^L [X_a^R\,, Q^R] + \frac{1}{2} (\lambda_2 - \lambda_2^\prime) i Q^L Q^R [X_a^R \,,Q^R] + i \frac{1}{2} \lambda_3^\prime \frac{1}{2} i Q^L \{ \widehat{X}_a^R \,, Q^R \} + \frac{1}{2} \lambda_4^\prime i Q^L \widehat{\omega}_a^R \,.
\end{multline}
Here $\chi_i$, $\chi_i^\prime$, $\lambda_i$ and $\lambda_i^\prime$ $i= (1,2,3,4)$ are Hermitian scalar fields over ${\cal M}$, the curly brackets denote anti-commutators throughout, and we have used
\beqa
&&\widehat{X}_a^L := \frac{1}{\ell_L+1/2} X_a^L \,, \quad {\widehat \omega}_a^L := \frac{1}{\ell_L +1/2} \omega_a^L \,, \nn \\
&& \widehat{X}_a^R := \frac{1}{\ell_R+1/2} X_a^R \,, \quad {\widehat \omega}_a^R := \frac{1}{\ell_R+1/2} \omega_a^R \,.
\eeqa
Let us also introduce the notation
\beqa
A_a^L &: =& {\tilde A}_a^L + i Q^R {\tilde A}_a^{\prime L} \nn \\
A_a^R &: =& {\tilde A}_a^R + i Q^L {\tilde A}_a^{\prime R}
\label{eq:newnot}
\eeqa
for future convenience.

\section{Reduction of the Yang-Mills Action over $S_F^2$}
\label{sec4}

Using the ${\rm SU(2)} \times {\rm SU(2)}$-equivariant gauge field in the action functional of the $U(4)$ Yang-Mills theory on ${\cal M} \otimes S_F^2 \times S_F^2 $, we can explicitly trace it  over the fuzzy spheres to reduce it to a theory on ${\cal M}$. It is quite useful to note the following identities
\begin{gather}
\label{identity1}
\lbrace Q \,, \lbrack X_a \,, Q \rbrack \rbrace = 0 \,, \quad  \lbrace X_a \,, \lbrack X_a \,, Q \rbrack \rbrace = 0 \,,
\quad (\mbox{sum over repeated $a$ is implied}) \,, \\
\label{identity2}
\lbrack Q \,, \lbrace X_a \,, Q \rbrace \rbrack = 0 \,, \quad  \lbrack X_a \,, \lbrace X_a \,, Q \rbrace \rbrack = 0 \,,
\quad (\mbox{sum over repeated $a$ is implied}) \,.
\end{gather}
which are valid for both the left and the right quantities and they significantly simplify the calculations, since they greatly reduce the number of traces to be computed.

The reduced action has the form
\be
S =   \int_{\cal M} {\cal L}_F + {\cal L}_G +  \frac{1}{g_L^2} V_1^L + \frac{1}{g_R^2} V_1^R + \frac{1}{g_{LR}^2} V_1^{L,R}
+ a_L^2 V^L_2 + a_R^2 V^R_2 \,.
\label{eq:reduced}
\ee
Each term in this expression is defined and evaluated below, while some details are relegated to the appendix C. 

\subsection{The Field Strength Term}

The field strength can be expressed as
\be
F_{\mu \nu} = \frac{1}{2} f_{\mu \nu}^L Q^L+ \frac{1}{2} f_{\mu \nu}^R Q^R + \frac{i}{2} g_{\mu \nu} {\mathbf 1}_4 + \frac{i}{2} h_{\mu \nu} Q^L Q^R
\ee
where
\beqa
f_{\mu \nu}^L &=& \partial_\mu a_\nu^L - \partial_\nu a_\mu^L \,, \quad f_{\mu \nu}^R = \partial_\mu a_\nu^R - \partial_\nu a_\mu^R \,, \nn \\
g_{\mu \nu} &=& \partial_\mu b_\nu - \partial_\nu b_\mu \,, \quad h_{\mu \nu} = \partial_\mu c_\nu - \partial_\nu c_\mu
\eeqa
The corresponding contribution to the Lagrangian is
\begin{eqnarray}
{\cal L}_F &:=& \frac{1}{4 g^2} \mbox{Tr}_{{\cal N}} \Big ( F_{\mu \nu}^\dagger F_{\mu \nu} \Big ) \nn \\
&=& \frac{1}{16 g^2}  \Big ( \left | f_{\mu\nu}^L \right |^2 + \left | f_{\mu\nu}^R \right |^2 + \left | g_{\mu\nu} \right |^2 + \left | h_{\mu\nu} \right |^2  
+ \frac{2}{(2 \ell_L + 1) (2 \ell_R + 1)}  f_{\mu\nu}^L  f_{\mu\nu}^R \nn \\
&&-\frac{1}{(2 \ell_R + 1)} \left ( f_{\mu\nu}^R g_{\mu \nu} -  f_{\mu\nu}^L h_{\mu \nu} \right ) 
-\frac{1}{(2 \ell_L + 1)} \left ( f_{\mu\nu}^L g_{\mu \nu} -  f_{\mu\nu}^R h_{\mu \nu} \right ) \nn \\
&& - \frac{2}{(2 \ell_L + 1) (2 \ell_R + 1)} g_{\mu \nu} h_{\mu \nu}
\Big ) \,.
\label{eq:strength}
\end{eqnarray}

\subsection{The Gradient Term}

The covariant derivatives are naturally expressed in two pieces
\begin{multline}
D_\mu \Phi_a^L = \frac{1}{2} \left (D_\mu (\chi_1 + \chi_1^\prime) + Q^L D_\mu (\chi_2 + \chi_2^\prime) \right ) [X_a^L\,, Q^L] + \frac{i}{4}\partial_{\mu}\chi_3 \{ \hat{X}_a^L, Q^L \} + \frac{1}{2} \partial_\mu \chi_4 \hat{\omega}_a^L \\
+i Q^R \left ( \frac{1}{2} \left (D_\mu (\chi_1 - \chi_1^\prime) + Q^L D_\mu (\chi_2 - \chi_2^\prime) \right ) [X_a^L\,, Q^L] + \frac{i}{4}\partial_{\mu}\chi_3^\prime \{ \hat{X}_a^L\,, Q^L \} + \frac{1}{2} \partial_\mu \chi_4^\prime \hat{\omega}_a^L \right )
\end{multline}
\begin{multline}
D_\mu \Phi_a^R = \frac{1}{2} \left (D_\mu (\lambda_1 + \lambda_1^\prime) + Q^R D_\mu (\lambda_2 + \lambda_2^\prime) \right ) [X_a^R\,, Q^R] + 
\frac{i}{4} \partial_{\mu} \lambda_3 \{ \hat{X}_a^R\,, Q^R \} + \frac{1}{2} \partial_\mu \lambda_4 \hat{\omega}_a^R \\
+i Q^L \left ( \frac{1}{2} \left (D_\mu (\lambda_1 - \lambda_1^\prime) + Q^R D_\mu (\lambda_2 - \lambda_2^\prime) \right ) [X_a^R\,, Q^R] + \frac{i}{4}\partial_{\mu}\lambda_3^\prime \{ \hat{X}_a^R\,, Q^R \} + \frac{1}{2} \partial_\mu \lambda_4^\prime \hat{\omega}_a^R \right )
\end{multline}
where we have $(i = 1,2)$
\beqa
D_\mu \chi_i &=& \partial_\mu \chi_i + \varepsilon_{ji} a_\mu \chi_j + \varepsilon_{ji} c_\mu \chi_j \nn \\
D_\mu \chi^\prime_i &=& \partial_\mu \chi_i^\prime + \varepsilon_{ji} a_\mu \chi_j^\prime - \varepsilon_{ji} c_\mu \chi_j^\prime \,.
\eeqa
with $(i = 1,2)$.

The gradient term takes the form 
\be
{\cal L}_G := {\cal L}_G^L + {\cal L}_G^R = {\mbox Tr}_{{\cal N}} \Big ( (D_\mu \Phi_a^L)^\dagger (D_\mu \Phi_a^L)  +  (D_\mu \Phi_a^R)^\dagger (D_\mu \Phi_a^R) \Big ) \,,
\ee
where
\beqa
&&{\cal L}_G^L = \frac{\ell_L (\ell_L +1)}{(\ell_L+1/2)^2} \Bigg [ \left ( 1 + \frac{1}{2 (\ell_R +1)} \right ) \left ((D_\mu \chi_1)^2 + (D_\mu \chi_2)^2 \right ) 
+ \left ( 1 - \frac{1}{2 (\ell_R +1)} \right ) \Big ( (D_\mu \chi_1^\prime)^2 \nn \\
&&+ (D_\mu \chi_2^\prime)^2 \Big ) \Bigg ] + \frac{1}{4} \frac{\ell_L (\ell_L +1)(\ell_L^2+\ell_L-1/4)}{(\ell_L+1/2)^4} \left \lbrack (\partial_\mu\chi_3)^2 + (\partial_\mu\chi_3^\prime)^2 
+ \frac{1}{(\ell_R +\frac{1}{2})} \partial_\mu\chi_3 \partial_\mu\chi_3^\prime \right \rbrack  \nn \\
\quad && + \frac{1}{2} \frac{\ell_L (\ell_L +1)}{(\ell_L+1/2)^3} \left \lbrack \partial_\mu \chi_3 \partial_\mu \chi_4 +  \partial_\mu \chi_3^\prime \partial_\mu \chi_4^\prime + \frac{1}{2 (\ell_R +1)} \left ( \partial_\mu \chi_3 \partial_\mu \chi_4^\prime + \partial_\mu \chi_3^\prime \partial_\mu \chi_4 \right )
 \right \rbrack \nn \\
 \quad&&+ \frac{1}{4}\frac{\ell_L^2+\ell_L+3/4}{(\ell_L+1/2)^2}  \left \lbrack (\partial_\mu\chi_4)^2 + (\partial_\mu\chi_4^\prime)^2 
+ \frac{1}{(\ell_R +\frac{1}{2})} \partial_\mu\chi_4 \partial_\mu\chi_4^\prime \right  \rbrack \,.
\eeqa
\beqa
&& {\cal L}_G^R = \frac{\ell_R (\ell_R+1)}{(\ell_R+1/2)^2} \Bigg \lbrack \left ( 1 + \frac{1}{2 (\ell_L +1)} \right ) \left( (D_\mu \lambda_1)^2 + (D_\mu \lambda_2)^2 \right ) + \left ( 1 - \frac{1}{2 (\ell_L +1)} \right ) \Big ( (D_\mu \lambda_1^\prime)^2  \nn \\
&& + (D_\mu \lambda_2^\prime)^2 \Big ) \Bigg \rbrack + \frac{1}{4} \frac{\ell_R (\ell_R +1)(\ell_R^2+\ell_R -1/4)}{(\ell_R+1/2)^4} \left \lbrack (\partial_\mu\lambda_3)^2 + (\partial_\mu\lambda_3^\prime)^2 + \frac{1}{(\ell_L +1)} \partial_\mu \lambda_3 \partial_\mu \lambda_3^\prime \right \rbrack \nn \\
\quad && + \frac{1}{2}  \frac{\ell_R (\ell_R +1)}{(\ell_R+1/2)^3} \left \lbrack \partial_\mu \lambda_3 \partial_\mu \lambda_4 + \partial_\mu \lambda_3^\prime 
\partial_\mu \lambda_4^\prime + \frac{1}{2 (\ell_L +1)} \left ( \partial_\mu \lambda_3 \partial_\mu \lambda_4^\prime + \partial_\mu \lambda_3^\prime \partial_\mu \lambda_4 \right ) \right \rbrack \nn \\
\quad&&+ \frac{1}{4}\frac{\ell_R^2+\ell_R+3/4}{(\ell_R+1/2)^2} \left \lbrack (\partial_\mu \lambda_4)^2 + (\partial_\mu \lambda_4^\prime)^2
+ \frac{1}{2 (\ell_L +1)} \partial_\mu \lambda_4 \partial_\mu \lambda_4^\prime  \right \rbrack \,.
\eeqa

It is useful to form the complex fields  
\be
\chi = \chi_1 + i \chi_2 \,, \quad {\bar \chi} = \chi_1 - i \chi_2 \,, \quad \lambda = \lambda_1 + i \lambda_2 \,, \quad {\bar \lambda} = \lambda_1 - i \lambda_2 \,,
\ee
then the covariant derivatives are expressed as 
\begin{gather}
D_\mu \chi = \partial_\mu \chi + i (a_\mu^L + c_\mu) \chi \,, \quad  D_\mu \chi^\prime = \partial_\mu \chi^\prime + i (a_\mu^L - c_\mu) \chi^\prime \,, \nn \\ 
D_\mu \lambda = \partial_\mu \lambda + i (a_\mu^R + c_\mu) \lambda \,, \quad  D_\mu \lambda^\prime = \partial_\mu \lambda^\prime + i (a_\mu^R - c_\mu) 
\lambda^\prime \,.
\end{gather}
We note that primed fields carry charge $-1$ under $c_\mu$. 

\subsection{The Potential Term}

Working with the duals, we have for $F_{ab}^L$
\be
\frac{1}{2} \varepsilon_{abc} F_{ab}^L  = \frac{1}{2} \epsilon_{abc} \lbrack \Phi_a^L,\Phi_b^L \rbrack -\Phi_c^L = F_c^L + i Q^R  {\tilde F}_c^L\,,
\ee
\begin{multline}
F_c^L = \frac{1}{2} \left (P_1^{L +} (\chi_1 + \chi_2 Q^L) + P_1^{L -} (\chi_1^\prime + \chi_2^\prime Q^L)  \right )  [X_c^L\,, Q^L] \\
+\frac{i}{4} \left ( 2 |\chi|^2 + 2 |\chi^\prime|^2  - P_2^L   \right ) \frac{ \{X_c^L \,,Q^L\} }{(\ell_L+1/2)} 
+ \frac{1}{4} P_3^L \frac{\omega_c^L}{(\ell_L+1/2)^2} \,,
\label{eq:dualcurvatureL1}
\end{multline}
\begin{multline}
{\tilde F}_c^L = \frac{1}{2} \left (P_1^{L +} (\chi_1 + \chi_2 Q^L) - P_1^{L -} (\chi_1^\prime + \chi_2^\prime Q^L)  \right )  [X_c^L\,, Q^L] \\
+ \frac{i}{4} \left ( 2 |\chi|^2 - 2 |\chi^\prime|^2 -  {\tilde P}_2^L  \right ) \frac{ \{X_c^L \,,Q^L\} }{(\ell_L+1/2)}
+ \frac{1}{4}  {\tilde P}_3^L \frac{\omega_c^L}{(\ell_L+1/2)^2} \,,
\label{eq:dualcurvatureL2}
\end{multline}
and $P_1^{L \pm}$, $P_2^L$ and $P_3^L$, $ {\tilde P}_2^L$, $ {\tilde P}_3^L$  are given in the appendix C.

Similarly for $F_{ab}^R$ we have
\be
\frac{1}{2} \varepsilon_{abc} F_{ab}^R = \frac{1}{2} \epsilon_{abc} \lbrack \Phi_a^R,\Phi_b^R \rbrack -\Phi_c^R = F_c^R + i Q^L  {\tilde F}_c^R\,,
\ee
\begin{multline}
F_c^R = \frac{1}{2} \left (P_1^{R +} (\lambda_1 + \lambda_2 Q^R) + P_1^{R -} (\lambda_1^\prime + \lambda_2^\prime Q^R)  \right )  [X_c^R\,, Q^R] \\
+\frac{i}{4} \left ( 2 |\lambda|^2 + 2 |\lambda^\prime|^2  - P_2^R \right ) \frac{ \{X_c^R \,,Q^R\}}{(\ell_R+1/2)} + \frac{1}{4} P_3^R \frac{\omega_c^L}{(\ell_R+1/2)^2} \,,
\label{eq:dualcurvatureR1}
\end{multline}
\begin{multline}
{\tilde F}_c^R = \frac{1}{2} \left (P_1^{R +} (\lambda_1 + \lambda_2 Q^R) - P_1^{R -} (\lambda_1^\prime + \lambda_2^\prime Q^R)  \right )  [X_c^R\,, Q^R] \\
+\frac{i}{4} \left ( 2 |\lambda|^2 - 2 |\lambda^\prime|^2  -  {\tilde P}_2^R  \right ) \frac{ \{X_c^R \,,Q^R\}}{(\ell_R+1/2)} + \frac{1}{4}  {\tilde P}_3^R \frac{\omega_c^L}{(\ell_R+1/2)^2} \,,
\label{eq:dualcurvatureR2}
\end{multline}
and $P_1^{R \pm}$, $P_2^R$ and $P_3^R$, $ {\tilde P}_2^R$, $ {\tilde P}_3^R$ are given in the appendix C.

In addition, we have for $F_{ab}^{L \,, R}$
\begin{multline}
F_{ab}^{L \,, R} = i \left ( (\chi_2 +\chi_2^\prime) - (\chi_1 +\chi_1^\prime) Q^L \right ) [X_a^L\,, Q^L] {\tilde A}_b^{\prime R} \\
+ i {\tilde A}_a^{\prime L}  \left ( (\lambda_2 +\lambda_2^\prime) - (\lambda_1 +\lambda_1^\prime) Q^R\right ) [X_b^R\,, Q^R] \,.
\end{multline}
where the notation introduced earlier in (\ref{eq:newnot}) is used.

With these we find for $V_1^L$, $V_1^R$ and $V_1^{L,R}$
\beqa
V_1^L &=& \mbox{Tr}_{{\cal N}} F_{ab}^{L  \dagger} F_{ab}^L \nn \\ 
&=& - 2 \mbox{Tr}_{{\cal N}} \left ( ( F_c^L)^2 + ( {\tilde F}_c^L)^2 + 2 i Q_R F_c^L {\tilde F}_c^L \right ) \nn \\
&=& T_1^L ( |\chi|^4 +  |\chi^\prime|^4 ) + T_2^L |\chi|^2 + {\tilde T}_2^L |\chi^\prime |^2 + T_3^L \,,
\eeqa

\beqa
V_1^R &=& \mbox{Tr}_{{\cal N}} F_{ab}^{R  \dagger} F_{ab}^R \nn \\ 
&=& - 2 \mbox{Tr}_{{\cal N}} \left ( ( F_c^R)^2 + ( {\tilde F}_c^R)^2 + 2 i Q_L F_c^R {\tilde F}_c^R \right ) \nn \\
&=& T_1^R ( |\lambda|^4 +  |\lambda^\prime|^4 ) + T_2^R |\lambda|^2 + {\tilde T}_2^R |\lambda^\prime |^2 + T_3^R \,,
\eeqa

\begin{multline}
V_1^{L,R} = 2 S_1 \left ( | \chi \lambda^\prime - \chi^\prime \lambda|^2 + | {\bar \lambda} \chi - \chi^\prime {\bar \lambda}^\prime |^2 \right ) + |\chi + \chi^\prime|^2 \left ( S_2^L \lambda_3^{\prime 2} + {\tilde S}_2^L \lambda_4^{\prime 2} + S_3^L \lambda_3^\prime \lambda_4^\prime \right ) \\
+  |\lambda+ \lambda^\prime|^2 \left ( S_2^R \chi_3^{\prime 2} + {\tilde S}_2^R \chi_4^{\prime 2} + S_3^R \chi_3^\prime \chi_4^\prime \right ) \,,
\end{multline}
where $T_1^{L,R}, T_2^{L,R}, {\tilde T}_2^{L,R}, T_3^{L,R}, S_1, S_2^L,{\tilde S}_2^L,S_3^L,S_2^R, {\tilde S}_2^R$ and $S_3^R$ are given in appendix C. 

\subsection{The Constraint Term}

Taking ${\tilde b}_L = \ell_L (\ell_L +1)$ and ${\tilde b}_R = \ell_R (\ell_R +1)$ as discussed earlier in section 2 we find
\begin{equation}
\label{constraint1}
\Phi_a^L \Phi_a^L + \ell_L(\ell_L+1) = R_1^L +  i Q^L R_2^L + i Q^R ( {\tilde R}_1^L +  i Q^L {\tilde R}_2^L ) \,,
\end{equation}
\begin{equation}
\label{constraint2}
\Phi_a^R \Phi_a^R + \ell_R(\ell_R+1) = R_1^R + i Q^R R_2^R + i Q^L ( {\tilde R}_1^R +  i Q^L {\tilde R}_2^R ) 
\end{equation}
where $R_1^L$, $R_2^L$ and ${\tilde R}_1^L$, ${\tilde R}_2^L$  and $R_1^R$, $R_2^R$ and ${\tilde R}_1^R$, ${\tilde R}_2^R$ are given in the appendix C.

The constraint terms in the action take the form 
\begin{multline}
V_2^L = (R_1^L)^2 + (R_2^L)^2 + ({\tilde R}_1^L)^2+({\tilde R}_2^L)^2 + \frac{1}{ (\ell_L + \frac{1}{2})} \left ( R_1^L  R_2^ L 
+ {\tilde R}_1^L {\tilde R}_2^L \right ) \\
+ \frac{1}{ (\ell_R + \frac{1}{2})} \left ( R_1^L  {\tilde R}_1^L + R_2^L {\tilde R}_2^L \right ) +\frac{1}{2 (\ell_L + \frac{1}{2}) (\ell_R + \frac{1}{2})} \left ( R_1^L  {\tilde R}_2^L + {\tilde R}_1^L R_2^L \right ) \,.
\end{multline}

\begin{multline}
V_2^R = (R_1^R)^2 + (R_2^R)^2 + ({\tilde R}_1^R)^2+({\tilde R}_2^R)^2 + \frac{1}{ (\ell_R + \frac{1}{2})} \left ( R_1^R  R_2^R 
+ {\tilde R}_1^R {\tilde R}_2^R \right ) \\
+ \frac{1}{ (\ell_L + \frac{1}{2})} \left ( R_1^R  {\tilde R}_1^R + R_2^R {\tilde R}_2^R \right )
+ \frac{1}{2 (\ell_L + \frac{1}{2}) (\ell_R + \frac{1}{2})} \left ( R_1^R  {\tilde R}_2^R 
+ {\tilde R}_1^R R_2^R \right )   \,.
\end{multline}

\subsection{Structure of the Reduced Theory}

In order to understand the structure of the reduced theory it is useful to analyze its vacuum structure. The potential has the form
\be
V = \frac{1}{g_L^2} V_1^L + \frac{1}{g_R^2} V_1^R + \frac{1}{g_{LR}^2} V_1^{L,R} + a_L^2 V^L_2 + a_R^2 V^R_2 \,.
\ee
Apart from the case $a_L= a_R = 0$, $V$ is zero if and only if $V_1^L, V_1^R, V_1^{L,R}, V^L_2, V^R_2$ all vanish.
Noting that zeros of $V_1^L, V_1^R, V_1^{L,R}$ coincide with zeros of the curvature terms, it is left to find the solutions of 
\be
F_{ab}^L = 0 \,, \quad F_{ab}^R = 0 \,, \quad F_{ab}^{L \,, R}=0 \,,
\ee
using the results obtained in the previous section. 

It turns out that the only solution to these equations, which is also a zero of both $V^L_2, V^R_2$ is given as
\beqa
&& |\chi| = |\chi^\prime| =  |\lambda| = |\lambda^\prime| = \frac{1}{2} \,, \nn \\
&&\chi \lambda^\prime = \chi^\prime \lambda\,, \quad {\bar \lambda} \chi = \chi^\prime {\bar \lambda}^\prime \,, \nn \\
&&\chi_3 = \chi_3^\prime = \chi_4 = \chi_4^\prime = 0 \,, \lambda_3 = \lambda_3^\prime = \lambda_4 = \lambda_4^\prime = 0 \,.
\label{eq:vacua}
\eeqa

In fact, the first condition on the second line together with the first line implies the second condition on the second line. It should be clear that vacua is not simply connected.
The first two lines of (\ref{eq:vacua}) imply that one of the complex fields can be written in terms of the other three. For instance, $\lambda^\prime = \frac{\chi^\prime \lambda}{\chi} = 4 \chi^\prime \lambda \bar{\chi}$. The vacuum manifold has therefore the structure of $T^3 = S^1 \times S^1\times S^1$, which has in particular
$\pi_1(T^3) = {\mathbb Z} \oplus {\mathbb Z} \oplus {\mathbb Z}$. 

Let us record the form of the action in the limit $\ell_L \,, \ell_R \rightarrow \infty$ which is going to be of essential interest in the next section.
\be
{\cal L}_F = \frac{1}{16 g^2} \Big ( \left | f_{\mu\nu}^L \right |^2 + \left | f_{\mu\nu}^R \right |^2 + \left | g_{\mu\nu} \right |^2 + \left | h_{\mu\nu} \right |^2 ) 
\ee

\begin{multline}
{\cal L}_G = |D_\mu \chi|^2 + |D_\mu \chi^\prime|^2 + |D_\mu \lambda|^2 + |D_\mu \lambda^\prime|^2 + \frac{1}{4} \Big( (\partial_\mu \chi_3)^2 + 
(\partial_\mu \chi_3^\prime)^2 + (\partial_\mu \chi_4)^2 + (\partial_\mu \chi_4^\prime)^2 \\
+ (\partial_\mu \lambda_3)^2 + (\partial_\mu \lambda_3^\prime)^2 + (\partial_\mu \lambda_4)^2 + (\partial_\mu \lambda_4^\prime)^2 \Big) \,.
\label{eq:LG}
\end{multline}

\begin{multline}
V_1^L = \frac{1}{g_L^2} \Bigg ( 4 \left( |\chi|^2 + \frac{1}{4}(\chi_3 + \chi_3^\prime) - \frac{1}{4} \right )^2  +  4 \left( |\chi^\prime|^2 + \frac{1}{4}(\chi_3 - \chi_3^\prime) - \frac{1}{4} \right )^2  \\
+ 2 (\chi_3 + \chi_3^\prime)^2 |\chi|^2 + 2 (\chi_3 - \chi_3^\prime)^2 |\chi^\prime|^2 + \frac{1}{2} (\chi_4^2 + \chi_4^{\prime 2}) \Bigg) \,.
\label{eq:V1L}
\end{multline}

\begin{multline}
V_1^R = \frac{1}{g_R^2} \Bigg ( 4 \left( |\lambda|^2 + \frac{1}{4}(\lambda_3 + \lambda_3^\prime) - \frac{1}{4} \right )^2  +  4 \left( |\lambda^\prime|^2 + \frac{1}{4}(\lambda_3 - \lambda_3^\prime) - \frac{1}{4} \right )^2  \\
+ 2 (\lambda_3 + \lambda_3^\prime)^2 |\lambda|^2 + 2 (\lambda_3 - \lambda_3^\prime)^2 |\lambda^\prime|^2 + \frac{1}{2} (\lambda_4^2 + \lambda_4^{\prime 2}) \Bigg) \,.
\label{eq:V1R}
\end{multline}

\begin{multline}
V_1^{L,R} \underset{\ell_L\,, \ell_R \rightarrow \infty}{=} \frac{1}{g_{L,R}^2} \Big ( 2 ( | \chi \lambda^\prime - \chi^\prime \lambda|^2 + | {\bar \lambda} \chi - \chi^\prime {\bar \lambda}^\prime |^2) -\frac{1}{2} \big ( |\chi + \chi^\prime|^2 (\lambda_3^{\prime 2} + \lambda_4^{\prime 2} ) \\
+ |\lambda+ \lambda^\prime|^2 ( \chi_3^{\prime 2} + \chi_4^{\prime 2} \big) \Big) \,.
\label{eq:VLR}
\end{multline}
 
\section{Vortices}
\label{sec6}

We will now discuss the vortex solutions of the reduced theory in the $\ell_L \,, \ell_R \rightarrow \infty$ limit. For simplicity, we restrict our attention to the case ${\cal M}=\mathbb{R}^2$. There is no canonical choice for the coefficients $a^2_L \,, a^2_R$ of the fuzzy constraint term; here we consider only the extreme cases of $a^2_L= a^2_R= 0$ and $a^2_L \,, a^2_R \rightarrow \infty$, which correspond respectively to imposing no constraint at all, and to imposing the constraints ``by hand''.  

\subsection{Case 1: No constraint}

As the constraint terms are absent, it is observed from the equations (\ref{eq:LG} - \ref{eq:VLR}) that $b_\mu$, $\chi_4$ and $\lambda_4$ decouple. In this case we have a $U(1)^3$ gauge theory. The vacuum has the nontrivial structure given in (\ref{eq:vacua}). On $\mathbb{R}^2$ this leads to vortices since the mapping of the circle at spatial infinity to the vacuum manifold 
\be
S^1(\infty) \longrightarrow T^3
\ee
is characterized by $\pi_1(T^3) = {\mathbb Z} \oplus {\mathbb Z} \oplus {\mathbb Z}$. 

To obtain a detailed description of these vortices we can select the radial gauge in which $a_r^L = a_r^R = c_r = 0$ and make the rotationally symmetric ansatz by setting
\beqa
&&\chi = \chi(r) e^{i n_1 \theta} \underset{r \rightarrow \infty}{\longrightarrow} \frac{1}{2} e^{i n_1 \theta} 
 \,, \quad \chi^\prime = \chi^\prime(r) e^{i n_2 \theta} \underset{r \rightarrow \infty}{\longrightarrow} \frac{1}{2} e^{i n_2 \theta}  \,, \nn \\
&&\lambda = \lambda(r) e^{i m_1 \theta} \underset{r \rightarrow \infty}{\longrightarrow} \frac{1}{2} e^{i m_1 \theta}
\,, \quad  \lambda^\prime = \lambda^\prime(r) e^{i m_2 \theta} \underset{r \rightarrow \infty}{\longrightarrow} \frac{1}{2} e^{i m_2 \theta} \,.  
\label{eq:astfields}
\eeqa
From (\ref{eq:vacua}) and (\ref{eq:astfields}) we see that the integers $n_1, n_2, m_1, m_2$ are not all independent but related to each other as
\be
(n_1 - n_2) - (m_1 - m_2) =  0 \,,
\label{eq:windingcond}
\ee
which is consistent with the fact that $\pi_1(T^3) = {\mathbb Z} \oplus {\mathbb Z} \oplus {\mathbb Z}$. In what follows we eliminate $m_2$ using (\ref{eq:windingcond})
and take the winding numbers of the complex fields as the set $(n_1, n_2, m_1)$. 

The real scalars are
\be
\chi_3 = \chi_3(r) \,, \quad \chi _3^\prime = \chi_3^\prime(r) \,, \quad \lambda_3 = \lambda_3(r) \,, \quad \lambda_3^\prime = \lambda_3^\prime(r) \,, \quad 
\chi _4^\prime = \chi_4^\prime(r) \,, \quad \lambda_4^\prime = \lambda_4^\prime(r) \,.
\label{eq:realscalars}
\ee
and they all tend to zero at spatial infinity ($r \rightarrow \infty$).

As for the gauge fields we have
\beqa
&& a_\theta^L =  a_\theta^L (r) \underset{r \rightarrow \infty}{\longrightarrow} - \frac{n_1+ n_2}{2} \,, \nn \\
&&a_\theta^R =  a_\theta^R (r) \underset{r \rightarrow \infty}{\longrightarrow} - \frac{m_1+ m_2}{2} = - \frac{2 m_1- (n_1- n_2)}{2}    \,, \nn \\
&&c_\theta = c_\theta(r)  \underset{r \rightarrow \infty}{\longrightarrow} - \frac{n_1 - n_2}{2} \,.
\eeqa

Asymptotic profiles of the fields listed above are all dictated by the finiteness of the action (\ref{eq:actionvortex}).

The action takes the form
\begin{multline}
S = 2 \pi \int _0^\infty r dr \Bigg [ \frac{1}{8 g^2} \Big ( \frac{1}{r^2} (\partial_r a_\theta^L)^2 + \frac{1}{r^2} (\partial_r a_\theta^R)^2 + \frac{1}{r^2} (\partial_r c_\theta)^2
\Big ) + (\partial_r \chi)^2 + \frac{1}{r^2} (n_1 + a_\theta^L + c_\theta)^2 \chi^2 \\
+ (\partial_r \chi^\prime)^2 + \frac{1}{r^2} (n_2 + a_\theta^L - c_\theta)^2 \chi^{\prime 2} + (\partial_r \lambda)^2 + \frac{1}{r^2} (m_1 + a_\theta^R + c_\theta)^2 \lambda^2 
+(\partial_r \lambda^\prime)^2 \\
+ \frac{1}{r^2} (m_1- (n_1- n_2) + a_\theta^R - c_\theta)^2 \lambda^{\prime 2} + \frac{1}{4} \Big ( (\partial_r \chi_3)^2 + (\partial_r \chi_3^\prime)^2 + (\partial_r \chi_4^\prime)^2  + (\partial_r \lambda_3)^2 \\
+ (\partial_r \lambda_3^\prime)^2  + (\partial_r \lambda_4^\prime)^2 \Big ) +  \frac{1}{g_L^2} \Bigg ( 4 \left( \chi^2 + \frac{1}{4}(\chi_3 + \chi_3^\prime) - \frac{1}{4} \right )^2  +  4 \left( \chi^{\prime 2} + \frac{1}{4}(\chi_3 - \chi_3^\prime) - \frac{1}{4} \right )^2  \\
+ 2 (\chi_3 + \chi_3^\prime)^2 \chi^2 + 2 (\chi_3 - \chi_3^\prime)^2 \chi^{\prime 2} + \frac{1}{2} \chi_4^{\prime 2} \Bigg) 
+ \frac{1}{g_R^2} \Bigg ( 4 \left( \lambda^2 + \frac{1}{4}(\lambda_3 + \lambda_3^\prime) - \frac{1}{4} \right )^2  \\
+  4 \left( \lambda^{\prime 2} + \frac{1}{4}(\lambda_3 - \lambda_3^\prime) - \frac{1}{4} \right )^2  + 2 (\lambda_3 + \lambda_3^\prime)^2 \lambda^2 + 2 (\lambda_3 - \lambda_3^\prime)^2 \lambda^{\prime 2} + \frac{1}{2} \lambda_4^{\prime 2} \Bigg) \\
+ \frac{1}{g_{L,R}^2} \Big ( 2 (\chi \lambda^\prime - \chi^\prime \lambda)^2 + 2 ({\bar \lambda} \chi - \chi^\prime {\bar \lambda}^\prime)^2 + F
\Bigg ]
\label{eq:actionvortex}
\end{multline}
where
\be
F = \left \lbrace
\begin{array}{ll}
-\frac{1}{2} \big ( (\chi + \chi^\prime)^2 (\lambda_3^{\prime 2} + \lambda_4^{\prime 2} ) + (\lambda+ \lambda^\prime)^2 ( \chi_3^{\prime 2} + \chi_4^{\prime 2} \big) \big)  & \mbox{for}  \quad n_1 = n_2 \\ [1.2 em]
-\frac{1}{2} \big ( (\chi^2 + \chi^{\prime 2}) (\lambda_3^{\prime 2} + \lambda_4^{\prime 2} ) + (\lambda^2+ \lambda^{\prime 2}) ( \chi_3^{\prime 2} + \chi_4^{\prime 2} \big) \big)  & \mbox{for} \quad n_1 \neq n_2
\end{array}
\right.
\label{eq:Fterm}
\ee

The equations of motion for the scalar and the gauge fields follow from (\ref{eq:actionvortex}) and (\ref{eq:Fterm}) in a straightforward manner. These are coupled non-linear differential equations for which we have not found any exact analytical solutions. However, it is possible to obtain the asymptotic profiles of the fields as $r \rightarrow \infty$. In this case we can write down the fluctuations around the vacuum values as
\begin{multline}
\chi = \frac{1}{2} - \delta \chi \,, \quad \chi^\prime = \frac{1}{2} - \delta \chi^\prime \,, \quad \lambda = \frac{1}{2} - \delta \lambda \,, \quad \lambda^\prime = \frac{1}{2} - \delta\lambda^\prime  \,, \quad a_\theta^L =  - \frac{n_1+ n_2}{2} + \delta a^L \,, \\[1.2 em]
\quad a_\theta^R =  - \frac{2 m_1- (n_1- n_2)}{2} + \delta a^R \,, \quad  c_\theta =  - \frac{n_1 - n_2}{2} + \delta c_\theta \,,
\end{multline}
while we can keep the same notation for the real scalars as they all fluctuate about the zero vacuum values. Assuming further that $(\frac{\delta a^L}{r})^2, 
(\frac{\delta a^R}{r})^2, (\frac{\delta C}{r})^2 $ are subleading\footnote{The region of validity of this approximation in terms of the parameters of the model will be given a little later on.} to the fluctuations in the complex and the real scalar fields, we obtain the following coupled set of 
linear second order differential equations:
\beqa
&& \partial_r^2 \delta a^L - \frac{1}{r} \delta a^L - 4 g^2 \delta a^L = 0 \,, \nn \\
&& \partial_r^2 \delta a^R - \frac{1}{r} \delta a^R - 4 g^2 \delta a^R = 0 \,, \nn \\
&& \partial_r^2 \delta c - \frac{1}{r} \delta c - 8 g^2 \delta c = 0 \,, \nn  \\
&& \partial_r^2 \delta \chi + \frac{1}{r} \delta \chi + \frac{4}{g_L^2} \left ( -\delta \chi + \frac{1}{4} (\chi_3 + \chi_3^\prime) \right) - \frac{1}{g_{L,R}^2} (\delta \chi - \delta \chi^\prime) = 0 \,, \nn \\
&& \partial_r^2 \delta \chi^\prime + \frac{1}{r} \delta \chi^\prime + \frac{4}{g_L^2} \left ( -\delta \chi + \frac{1}{4} (\chi_3 - \chi_3^\prime) \right) + \frac{1}{g_{L,R}^2} (\delta \chi - \delta \chi^\prime) = 0 \,, \nn  \\
&& \partial_r^2 \delta \lambda + \frac{1}{r} \delta \lambda + \frac{4}{g_R^2} \left ( -\delta \lambda + \frac{1}{4} (\lambda_3 + \lambda_3^\prime) \right) - \frac{1}{g_{L,R}^2} (\delta \lambda - \delta \lambda^\prime) = 0 \,, \nn  \\
&& \partial_r^2 \delta \lambda^\prime + \frac{1}{r} \delta \lambda^\prime + \frac{4}{g_R^2} \left ( -\delta \lambda + \frac{1}{4} (\lambda_3 - \lambda_3^\prime) \right) + \frac{1}{g_{L,R}^2} (\delta \lambda - \delta \lambda^\prime) = 0 \,,\nn  \\
&& \partial_r^2 \chi_3 + \frac{1}{r} \chi_3 + \frac{4}{g_L^2} \left ( \delta \chi + \delta \chi^\prime - \frac{3}{2} \chi_3 \right ) = 0 \,, \\
&& \partial_r^2 \chi_3^\prime + \frac{1}{r} \chi_3^\prime + \frac{4}{g_L^2} \left ( \delta \chi - \delta \chi^\prime - \frac{3}{2} \chi_3^\prime \right ) + \frac{\gamma}{g_{L,R}^2} \chi_3^\prime = 0 \,, \nn \\
&& \partial_r^2 \lambda_3 + \frac{1}{r} \lambda_3 + \frac{4}{g_R^2} \left ( \delta \lambda + \delta \lambda^\prime - \frac{3}{2} \lambda_3 \right ) = 0 \,, \nn  \\
&& \partial_r^2 \lambda_3^\prime + \frac{1}{r} \lambda_3^\prime + \frac{4}{g_R^2} \left ( \delta \lambda - \delta \lambda^\prime - \frac{3}{2} \lambda_3^\prime \right ) + \frac{\gamma}{g_{L,R}^2} \lambda_3^\prime = 0 \,, \nn \\
&& \partial_r^2 \chi_4^\prime + \frac{1}{r} \chi_4^\prime - \frac{2}{g_L^2} \chi_4^\prime + \frac{\gamma}{g_{L,R}^2} \chi_4^\prime = 0 \,, \nn \\
&& \partial_r^2 \lambda_4^\prime + \frac{1}{r} \lambda_4^\prime - \frac{2}{g_R^2} \lambda_4^\prime + \frac{\gamma}{g_{L,R}^2} \lambda_4^\prime = 0 \,. \nn 
\eeqa
where 
\be
\gamma = \left \lbrace
\begin{array}{ll}
1 \quad & \mbox{for} \quad n_1 \neq n_2 \\
2 \quad & \mbox{for} \quad n_1 = n_2 \,.
\end{array}
\right.
\ee
The gauge fields have the asymptotic profiles
\beqa
\delta a^L &=& F^L r K_1 (2 g r)  \nn \\
\delta a^R &=& F^R r K_1 (2 g r) \nn \\
\delta c &=& F r K_1 (2 \sqrt{2} g r) \,.
\eeqa
Some algebra yields the asymptotic profiles of the scalar fields as 
\beqa
\delta \chi &=& C_1 K_0 \left (\frac{\sqrt{2} r }{g_L} \right) + C_2 K_0 \left (\frac{2\sqrt{2} r }{g_L} \right) + C_3 K_0 \left (\sqrt{\alpha_+^L} r \right) + C_4 K_0 \left (\sqrt{\alpha_-^L} r \right) \nn \\
\delta \chi^\prime &=& C_1 K_0 \left (\frac{\sqrt{2} r }{g_L} \right) + C_2 K_0 \left (\frac{2\sqrt{2} r }{g_L} \right) - C_3 K_0 \left (\sqrt{\alpha_+^L} r \right) - C_4 K_0 \left (\sqrt{\alpha_-^L} r \right) \nn \\
\chi_3 &=& C_1 K_0 \left (\frac{\sqrt{2} r }{g_L} \right) -  2 C_2 K_0 \left (\frac{2\sqrt{2} r }{g_L} \right) \nn \\
\chi_3^\prime &=&  C_3^\prime K_0 \left (\sqrt{\alpha_+^L} r \right) + C_4^\prime K_0 \left (\sqrt{\alpha_-^L}  r \right) \nn \\ 
\chi_4^\prime &=&  C_5 K_0 \left (\sqrt{\beta^L} r \right) \nn \\
\delta \lambda &=& D_1 K_0 \left (\frac{\sqrt{2} r }{g_R} \right) + D_2 K_0 \left (\frac{2\sqrt{2} r }{g_R} \right) + D_3 K_0 \left (\sqrt{\alpha_+^R} r \right) + D_4 K_0 \left (\sqrt{\alpha_-^R} r \right) \nn \\
\delta \lambda^\prime &=& D_1 K_0 \left (\frac{\sqrt{2} r }{g_R} \right) + D_2 K_0 \left (\frac{2\sqrt{2} r }{g_R} \right) - D_3 K_0 \left (\sqrt{\alpha_+^R} r \right) - D_4 K_0 \left (\sqrt{\alpha_-^R} r \right) \nn \\
\lambda_3 &=& D_1 K_0 \left (\frac{\sqrt{2} r }{g_R} \right) -  2 D_2 K_0 \left (\frac{2\sqrt{2} r }{g_R} \right) \nn \\
\lambda_3^\prime &=&  D_3^\prime K_0 \left (\sqrt{\alpha_+^R} r \right) + D_4^\prime K_0 \left (\sqrt{\alpha_-^R} r \right) \nn \\ 
\lambda_4^\prime &=&  D_5 K_0 \left (\sqrt{\beta^R} r \right) 
\eeqa
where
\be
\left \lbrace
\begin{array}{ll}
\alpha_\pm^L = \frac{5}{g_L^2} + \frac{1}{2 g_{L,R}^2} \pm \frac{1}{2} \left (\frac{36}{g_L^4} - \frac{12}{g_L^2 g_{L,R}^2} + \frac{9}{g_{L,R}^4} \right )^{\frac{1}{2}}  & \mbox{for} \quad \gamma = 1 \\
\alpha_\pm^L = \frac{5}{g_L^2} \pm \left (\frac{9}{g_L^4} - \frac{4}{g_L^2 g_{L,R}^2} + \frac{4}{g_{L,R}^4} \right )^{\frac{1}{2}} & \mbox{for} \quad \gamma = 2 
\end{array}
\right.
\ee
\be
\left \lbrace
\begin{array}{ll}
\alpha_\pm^R = \frac{5}{g_R^2} + \frac{1}{2 g_{L,R}^2} \pm \frac{1}{2} \left (\frac{36}{g_R^4} - \frac{12}{g_R^2 g_{L,R}^2} + \frac{9}{g_{L,R}^4} \right )^{\frac{1}{2}}  & \mbox{for} \quad \gamma = 1 \\
\alpha_\pm^R =\frac{5}{g_R^2} \pm \left (\frac{36}{g_R^4} - \frac{4}{g_R^2 g_{L,R}^2} + \frac{4}{g_{L,R}^4} \right )^{\frac{1}{2}}  & \mbox{for} \quad \gamma = 2 
\end{array}
\right.
\ee
\be
\beta^L = \frac{2}{g_L^2} - \frac{\gamma}{g_{L,R}^2} \,, \quad \beta^R = \frac{2}{g_R^2} - \frac{\gamma}{ g_{L,R}^2} \,.
\ee
We further have that the coefficients $C_3^\prime$ and $C_4^\prime$ are fixed in terms of $C_3$ and $C_4$ as 
\beqa
&&\gamma = 1 \rightarrow 
\left \lbrace 
\begin{array}{l}
C_3^\prime = \left (\frac{g_L^2}{2} \alpha_+^L - \frac{g_L^2}{g_{L,R}^2} -2 \right ) C_3 \,, \\
C_4^\prime = \left (\frac{g_L^2}{2} \alpha_-^L - \frac{g_L^2}{g_{L,R}^2} -2 \right ) C_4 \,.
\end{array} 
\right. \\
&&\gamma = 2 \rightarrow 
\left \lbrace 
\begin{array}{l}
C_3^\prime = -2  C_3 \,, \\
C_4^\prime = C_4 \,,
\end{array} 
\right.
\eeqa
and likewise for the $D_3^\prime$ and $D_4^\prime$ 
\beqa
&&\gamma = 1 \rightarrow 
\left \lbrace 
\begin{array}{l}
D_3^\prime = \left (\frac{g_R^2}{2} \alpha_+^R - \frac{g_R^2}{g_{L,R}^2} -2 \right ) D_3 \,, \\
D_4^\prime = \left (\frac{g_R^2}{2} \alpha_-^R - \frac{g_R^2}{g_{L,R}^2} -2 \right ) D_4 \,.
\end{array} 
\right. \\
&&\gamma = 2 \rightarrow 
\left \lbrace 
\begin{array}{l}
D_3^\prime = -2  D_3 \,, \\
D_4^\prime = D_4 \,.
\end{array} 
\right.
\eeqa
The coefficients $C_a, D_a, F^L, F_R, F, (a= 1 \,, \cdots\,, 5 )$ can be found by numerical methods. Such a numerical computation was given in \cite{Seckin-Derek}, for the case of $U(1)$ vortices emerging from the equivariant reduction of a $U(2)$ theory over ${\cal M} \times S_F^2$. We will not go into numerical calculations in this article. 
However, we can still note a few qualitative features stemming from the asymptotic profiles of fields listed above. Focusing on the special case, $g_L = g_R = \sqrt{2} g_{L,R} : = {\tilde g}$, the expressions above simplify to
\be
\left \lbrace
\begin{array}{ll}
\alpha_\pm^L = \frac{6 \pm 2 \sqrt{3}}{{\tilde g}^2} & \mbox{for} \quad \gamma = 1\,, \\ [1em]
\alpha_\pm^L = \frac{6 \pm \sqrt{17}}{2 {\tilde g}^2} & \mbox{for} \quad \gamma = 2 \,,
\end{array}
\right.
\ee
and $\beta = \frac{2}{{\tilde g}^2}$ for $\gamma = 1$. For $\gamma = 2$, it is easily observed that there are no fluctuations in zero vacuum value of the fields $\chi_4^\prime$ and $\lambda_4^\prime$ at this approximation. It follows from the asymptotic form of the Bessel functions that, $(\frac{\delta a^L}{r})^2, (\frac{\delta a^R}{r})^2, (\frac{\delta C}{r})^2 $ are subleading to the fluctuations in the complex and the real scalar fields, as long as $4g > \frac{\sqrt{2}}{{\tilde g}}$. Furthermore, the field strengths decay faster than the scalar fields if $2g > \frac{\sqrt{2}}{{\tilde g}}$. This result indicates that vortices tend to attract as long as $2g > \frac{\sqrt{2}}{{\tilde g}}$, since it is known that field strengths are responsible for the repulsive and scalars are responsible for the attractive forces between vortices \cite{Vortices}. In particular, the reduced "standard" Yang-Mills theory with $g {\tilde g} = 1$ falls into this region of the parameter space.

\subsection{Case 2: The constraints fully imposed}

The fuzzy constraints
\be
\Phi_a^L \Phi_a^L + \ell_L(\ell_L+1) =0 \,, \quad  \Phi_a^R \Phi_a^R + \ell_R(\ell_R+1) = 0 
\ee
are equivalent to the algebraic equations
\be
R_1^L = 0 \,, \quad  R_2^L = 0 \,, \quad  {\tilde R}_1^L = 0 \,,  \quad {\tilde R}_2^L = 0 \,, \quad  R_1^R = 0 \,,  \quad R_2^R = 0 \,, \quad  {\tilde R}_1^R = 0 \,, \quad {\tilde R}_2^R = 0 \,.
\label{eq:algebraicconstraints}
\ee
where expressions for all $R$ are given in the appendix C. These equations can be solved order by order in powers of the parameters $\frac{1}{\ell_L}$ and $\frac{1}{\ell_R}$ to obtain expressions for the real scalar fields in terms of the modulus of the complex scalars in the theory. Substituting the leading order solutions of the real fields yields an action involving the complex scalars only.

To leading order in $\frac{1}{\ell_L}$ and $\frac{1}{\ell_R}$, (\ref{eq:algebraicconstraints}) yield
\beqa
&&\chi_3 = \frac{1}{\ell_L^2}  ( |\chi|^2 + |\chi^\prime|^2 - \frac{1}{2} ) \,, \quad \chi_4 = - \frac{1}{\ell_L} ( |\chi|^2 + |\chi^\prime|^2 - \frac{1}{2} ) \,,  \nn \\
&&\chi_3^\prime = \frac{1}{\ell_L^2} ( |\chi|^2 - |\chi^\prime|^2) \,, \quad \chi_4^\prime = -\frac{1}{\ell_L} ( |\chi|^2 - |\chi^\prime|^2 ) \,,  \nn \\
&&\lambda_3 = \frac{1}{\ell_R^2}  ( |\lambda|^2 + |\lambda^\prime|^2 - \frac{1}{2} ) \,, \quad \lambda_4 = - \frac{1}{\ell_R} ( |\lambda|^2 + |\lambda^\prime|^2 - \frac{1}{2} ) \,, \nn \\
&&\lambda_3^\prime = \frac{1}{\ell_R^2} ( |\lambda|^2 - |\lambda^\prime|^2) \,, \quad \lambda_4^\prime = -\frac{1}{\ell_R} ( |\lambda|^2 - |\lambda^\prime|^2 ) \,,
\label{eq:constraintsolutions}
\eeqa
Substituting (\ref{eq:constraintsolutions}) into the reduced action obtained in section 3 gives
\beqa
&&S = \int d^2 y \frac{1}{16 g^2} \Bigg ( \big ( 1 - \frac{1}{16 \ell_L^2}  \big ) | f_{\mu\nu}^L |^2 + \big ( 1 - \frac{1}{16 \ell_R^2}  \big ) | f_{\mu\nu}^R |^2
+ \frac{3}{8 \ell_L  \ell_R}  f_{\mu\nu}^L  f_{\mu\nu}^R +  | h_{\mu\nu} |^2 \nn \\
&&+ \frac{1}{2}\left(\frac{1}{\ell_R} - \frac{1}{\ell_R^2} \right) h_{\mu\nu}  f_{\mu\nu}^L \nn 
+ \frac{1}{2}\left(\frac{1}{\ell_L} - \frac{1}{\ell_L^2} \right) h_{\mu\nu}  f_{\mu\nu}^R \Bigg ) 
+ \left (1 - \frac{1}{4 \ell_L^2} + \frac{1}{2 (\ell_R +1)} \right ) |D_\mu \chi|^2 \nn \\
&&+ \left (1 - \frac{1}{4 \ell_L^2} - \frac{1}{2 (\ell_R +1)} \right ) |D_\mu \chi^\prime|^2 \nn + \left (1 - \frac{1}{4 \ell_R^2} + \frac{1}{2 (\ell_L +1)} \right ) |D_\mu \lambda|^2 \nn \\ 
&&+ \left (1 - \frac{1}{4 \ell_R^2} - \frac{1}{2 (\ell_L +1)} \right )|D_\mu \lambda^\prime|^2 +  \frac{1}{2 \ell_L^2} \left ( \left ( \partial_\mu |\chi|^2 \right )^2 +  \left ( \partial_\mu |\chi^\prime|^2 \right )^2  \right )  \\
&&+ \frac{1}{2 \ell_R^2} \left ( \left ( \partial_\mu |\lambda|^2 \right )^2 +  \left ( \partial_\mu |\lambda^\prime|^2 \right )^2 \right ) 
+ \frac{4}{g_L^2} \left ( 1 +  \frac{5}{4 \ell_L^2}  \right ) \left ( \left (  |\chi|^2 - \frac{1}{4} \right )^2 + \left ( |\chi^\prime|^2 - \frac{1}{4} \right )^2 \right )  \nn \\
&& +\frac{4}{g_R^2} \left ( 1 +  \frac{5}{4 \ell_R^2}  \right ) \left ( \left (  |\lambda|^2 - \frac{1}{4} \right )^2 + \left ( |\lambda^\prime|^2 - \frac{1}{4} \right )^2 \right )
+ \frac{1}{g_{L,R}^2} \Bigg ( 2 ( | \chi \lambda^\prime - \chi^\prime \lambda|^2 \nn \\
&&+ | {\bar \lambda} \chi - \chi^\prime {\bar \lambda}^\prime |^2) -\frac{1}{2 \ell_R^2} |\chi + \chi^\prime|^2 (|\lambda|^2 - |\lambda^\prime|^2)^2 - \frac{1}{2 \ell_L^2}  |\lambda+ \lambda^\prime|^2  (|\chi|^2 - |\chi^\prime|^2)^2  \Bigg )\,. \nn
\eeqa
where we have already solved the equations of motion for $b_\mu$ and inserted
\be
g_{\mu \nu} = \frac{1}{4}\left(\frac{1}{\ell_L} - \frac{1}{\ell_L^2} \right) f_{\mu \nu}^L +
\frac{1}{4}\left(\frac{1}{\ell_R} - \frac{1}{\ell_R^2} \right)f_{\mu \nu}^R + \frac{1}{4 \ell_L \ell_R} h_{\mu \nu} \,.
\ee 
It is readily observed that the minimum of the potential resides at
\beqa
&& |\chi| = |\chi^\prime| =  |\lambda| = |\lambda^\prime| = \frac{1}{2} \,, \nn \\
&&\chi \lambda^\prime = \chi^\prime \lambda\,, \quad {\bar \lambda} \chi = \chi^\prime {\bar \lambda}^\prime \,. 
\eeqa
We can again pick the radial gauge, and make the rotationally symmetric ansatz to look for vortex solutions. The action takes the form

\begin{multline}
S = 2 \pi \int _0^\infty r dr \Bigg [ \frac{1}{8 g^2} \Bigg ( \frac{1}{r^2} \big ( 1 - \frac{1}{16 \ell_L^2}  \big ) (\partial_r a_\theta^L)^2 + 
\big ( 1 - \frac{1}{16 \ell_R^2}  \big ) \frac{1}{r^2} (\partial_r a_\theta^R)^2 + \frac{1}{r^2} (\partial_r c_\theta)^2 \Big ) \\
+  \frac{1}{r^2}  \frac{3}{8 \ell_L  \ell_R}  (\partial_r a_\theta^L)  (\partial_r a_\theta^R) 
+ \frac{1}{r^2} \frac{1}{2}\left(\frac{1}{\ell_R} - \frac{1}{\ell_R^2} \right)  (\partial_r a_\theta^L) (\partial_r c_\theta) + \frac{1}{r^2} \frac{1}{2}\left(\frac{1}{\ell_L} - \frac{1}{\ell_L^2} \right) (\partial_r a_\theta^R) (\partial_r c_\theta)  \Bigg ) \\
+ \left (1 - \frac{1}{4 \ell_L^2} + \frac{1}{2 (\ell_R +1)} \right ) \left ( (\partial_r \chi)^2 + \frac{1}{r^2} (n_1 + a_\theta^L + c_\theta)^2 \chi^2 \right ) \\
+  \left (1 - \frac{1}{4 \ell_L^2} - \frac{1}{2 (\ell_R +1)} \right ) \left ( (\partial_r \chi^\prime)^2 + \frac{1}{r^2} (n_2 + a_\theta^L - c_\theta)^2 \chi^{\prime 2} \right ) \\
+ \left (1 - \frac{1}{4 \ell_R^2} + \frac{1}{2 (\ell_L +1)} \right ) \left ( (\partial_r \lambda)^2 + \frac{1}{r^2} (m_1 + a_\theta^R + c_\theta)^2 \lambda^2 \right ) \\
+ \left (1 - \frac{1}{4 \ell_R^2} - \frac{1}{2 (\ell_L +1)} \right ) \left ( (\partial_r \lambda^\prime)^2 + \frac{1}{r^2} (m_1- (n_1- n_2) + a_\theta^R - c_\theta)^2 \lambda^{\prime 2} \right )  \\
+ \frac{2}{\ell_L^2} \left ( \chi^2 (\partial_r \chi)^2 +   \chi^{\prime 2} (\partial_r \chi^\prime)^2 \right ) + \frac{2}{\ell_R^2} \left ( \lambda^2 (\partial_r \lambda)^2 +   \lambda^{\prime 2} (\partial_r \lambda^\prime)^2 \right ) \\
+ \frac{4}{g_L^2} \left ( 1 +  \frac{5}{4 \ell_L^2}  \right ) \left ( \left (  \chi^2 - \frac{1}{4} \right )^2 + \left ( \chi^{\prime 2} - \frac{1}{4} \right )^2 \right )
+\frac{4}{g_R^2} \left ( 1 +  \frac{5}{4 \ell_R^2}  \right ) \left ( \left (  \lambda^2 - \frac{1}{4} \right )^2 + \left ( \lambda^{\prime 2} - \frac{1}{4} \right )^2 \right ) \\
+ \frac{1}{g_{L,R}^2} \Bigg ( 2 ( \chi \lambda^\prime - \chi^\prime \lambda)^2 + 2 (\lambda \chi - \chi \lambda^\prime)^2 + F \Bigg ) \Bigg]
\end{multline}
where
\be
F = \left \lbrace
\begin{array}{ll}
-\frac{1}{2 \ell_R^2} (\chi + \chi^\prime)^2 (\lambda^2 - \lambda^{\prime 2})^2  -\frac{1}{2 \ell_L^2} (\lambda + \lambda^\prime)^2 (\chi^2 - \chi^{\prime 2})^2  & \mbox{for}  \quad n_1 = n_2 \\ [1.2 em]
-\frac{1}{2 \ell_R^2} (\chi^2 + \chi^{\prime 2}) (\lambda^2 - \lambda^{\prime 2})^2 -\frac{1}{2 \ell_L^2} (\lambda^2 + \lambda^{\prime 2}) (\chi^2 - \chi^{\prime 2})^2  & \mbox{for} \quad n_1 \neq n_2
\end{array}
\right.
\ee

To leading order asymptotic profiles of the gauge fields are 
\beqa
\delta a^L &=& \alpha_1 r K_1 (2 g r)  + \alpha_2 r K_1 \left ( 2 g \left (1 + \frac{1}{4} \left (\frac{1}{\ell_L^2} + \frac{1}{\ell_R^2} \right ) \right ) r \right ) \,, \nn \\
\delta a^R &=& \alpha_1 r K_1 (2 g r)  + \alpha_2 r K_1 \left ( 2 g \left (1 + \frac{1}{4} \left (\frac{1}{\ell_L^2} + \frac{1}{\ell_R^2} \right ) \right ) r \right ) \,,\nn \\
\delta c &=& \alpha_3 r K_1 \left ( 2\sqrt{2} g \left (1 - \frac{3}{8} \left (\frac{1}{\ell_L^2} + \frac{1}{\ell_R^2} \right ) \right ) r \right ) \,.
\eeqa

The asymptotic profiles of the scalar fields read
\beqa 
\delta \chi &=& C_1 K_0 \left (\sqrt{\mu_1} r \right) + C_2 K_0 \left (\sqrt{\mu_2} r \right) \nn \\
\delta \chi^\prime &=& C_1^\prime K_0 \left (\sqrt{\mu_1} r \right) + C_2^\prime K_0 \left (\sqrt{\mu_2} r \right) \nn \\
\delta \lambda &=& C_3 K_0 \left (\sqrt{\nu_1} r \right) + C_4 K_0 \left (\sqrt{\nu_2} r \right) \nn \\
\delta \chi &=& C_3^\prime K_0 \left (\sqrt{\nu_1} r \right) + C_4^\prime K_0 \left (\sqrt{\nu_2} r \right) 
\eeqa
Focusing on the case $g_L = g_R = {\sqrt2} g_{L,R} : = {\tilde g}$, we find
\beqa
\sqrt{\mu_1} &=& \frac{2 \sqrt{2}}{{\tilde g}} \left ( 1 + \frac{1}{4 \ell_L^2} \right) \,, \nn \\
\sqrt{\mu_2} &=& \frac{2}{{\tilde g}} \left ( 1 + \frac{3}{8 \ell_L^2} - \frac{3}{8 \ell_R^2} \right) \,, \nn \\
\sqrt{\nu_1} &=& \frac{2 \sqrt{2}}{{\tilde g}} \left ( 1 + \frac{1}{4\ell_R^2} \right ) \,, \nn \\
\sqrt{\nu_2} &=& \frac{2}{{\tilde g}} \left ( 1 + \frac{3}{8 \ell_R^2} - \frac{3}{8 \ell_L^2} \right) \,.
\eeqa
In this case, it follows from the asymptotic form of the Bessel functions that, $(\frac{\delta a^L}{r})^2, (\frac{\delta a^R}{r})^2$, $(\frac{\delta C}{r})^2 $ are subleading to the fluctuations in the complex and the real scalar fields, as long as $4g > \frac{2 \sqrt{2}}{{\tilde g}}$. For finite values of $\ell_L \,, \ell_R$, at the critical $g {\tilde g} = 1$ coupling the vortices tend to repel since the scalars decay faster than the field strength. In particular, in the strict limit $\ell_L \,, \ell_R \rightarrow \infty$ the model collapses to the critically coupled BPS vortices at $g {\tilde g} = 1$. The BPS bound for this model can be written. 
Saturating the bound gives the action
\beqa
S &=& \frac{\pi}{2} (n_1 + n_2 + m_1 + m_2) \nn \\
&=& \pi (n_2 + m_1) \,,
\eeqa
since $m_2 = - (n_1-n_2) + m_1$ and the BPS equations are
\beqa
D_1 \chi \pm i D_2 \chi &=& 0 \,, \quad D_1 \chi^\prime \pm i D_2 \chi^\prime = 0 \,, \nn \\
D_1 \lambda \pm i D_2 \lambda &=& 0 \,, \quad D_1 \lambda^\prime \pm i D_2 \lambda^\prime = 0 \,.
\eeqa
\begin{multline}
B^L + \frac{1}{\sqrt{2}} B \mp 4 \sqrt{2} g^2 \left (|\chi|^2 - \frac{1}{4} \right ) = 0 \,, \quad 
B^L - \frac{1}{\sqrt{2}} B \mp 4 \sqrt{2} g^2 \left (|\chi^\prime|^2 - \frac{1}{4} \right ) = 0 \,, \\
\noindent B^R + \frac{1}{\sqrt{2}} B \mp 4 \sqrt{2} g^2 \left ((|\lambda|^2 - \frac{1}{4} \right ) = 0 \,, \quad
B^R - \frac{1}{\sqrt{2}} B \mp 4 \sqrt{2} g^2 \left (|\lambda^\prime|^2 - \frac{1}{4} \right ) = 0 \,,
\end{multline}
together with the supplementary conditions  
\be
\chi \lambda^\prime = \chi^\prime \lambda\,, \quad {\bar \lambda} \chi = \chi^\prime {\bar \lambda}^\prime \,,
\ee
and where $B^L = f^L_{r \theta}$, $B^R = f^R_{r \theta}$, $B = h^L_{r \theta}$. A similar model, though on the noncommutative plane ${\mathbb R}_\theta^2$ have appeared in \cite{Lechtenfeld1}. We have not found any reference in the literature studying the solutions of these BPS equations however we think that, in principal, it may be possible to construct them using the methods of \cite{Taubes, Vortices}. This is beyond the scope of the present article.

\section{Conclusions}

In the present article, we have investigated the ${\rm SU(2)} \times {\rm SU(2)} $ equivariant reduction of a $U(4)$ gauge theory over $S_F^2 \times S_F^2$. We have started from an $SU({\cal N})$ gauge theory suitably coupled to a set of scalar fields in the adjoint of $SU({\cal N})$ on a manifold ${\cal M}$, which leads in general to a $U(n)$ gauge theory on 
${\cal M} \times S_F^2 \times S_F^2 $  after spontaneous symmetry breaking. Focusing on the $U(4)$ theory we have determined the most general ${\rm SU(2)} \times {\rm SU(2)} $-equivariant $U(4)$ gauge fields and performed the dimensional reduction of  the theory over $S_F^2 \times S_F^2$.  We have found that the emergent model is a $U(1)^4$ gauge theory coupled to four complex and eight real scalar fields. Studying  this theory on ${\mathbb R}^{2}$ in two different limiting cases we have demonstrated that, these particular models have vortex solutions with $U(1)^3$ gauge symmetry which tend to attract or repel at the critical point of the parameter space $g {\tilde g} = 1$ as discussed in the previous section.

We find this line of research very interesting as it gives us concrete results on the structure of gauge theories with fuzzy extra dimensions.In particular, we are interested in investigating the $SU(2)$-equivariant formulation of a $U(3)$ gauge theory on ${\cal M} \times S_F^2$. In this case, $SU(2)$ gauge transformations in $U(3)$ are generated by the $SU(2)$ rank $1$ and rank $2$ irreducible tensors in the adjoint representation of $SU(2)$ and among the rotational invariants of the symmetry generators, suitably contracted rank two tensor operators over the fuzzy sphere also appear. In other words, and somewhat more accurately, fuzzy version of $x_a x_b Q_{ab}$, $Q_{ab}$ being the quadrupole tensor carrying the spin $2$representations of $SU(2)$, appears as another rotational invariant in the theory whose contribution should be taken into account. We will report on these and related developments elsewhere in the near future.

\vskip 2em

{\bf Acknowledgements}

\vskip 1em
I thank A.P. Balachandran and S. Vaidya for useful discussions. I also thank A. Behtash for proofreading the article. This work is supported by T\"{U}BiTAK under project No. 110T738, T\"{U}BA-GEBiP program of The Turkish Academy of Sciences and the Middle East Technical University under Project No. BAP- 08-11-2010-R-108.

\appendices

\subsection{$S_F^2$ and $S_F^2 \times S_F^2$}

The fuzzy sphere at level $\ell$ is defined to be the algebra of $(2\ell +1) \times (2 \ell +1)$ matrices $\mbox{Mat}(2 \ell +1)$.  The three Hermitian ``coordinate functions''
\be
{\hat x}_a := \frac{i}{\sqrt{\ell (\ell +1)}} X_a^{(2\ell+1)}
\ee
satisfy
\be
\lbrack {\hat x}_a \,, {\hat x}_b \rbrack = \frac{i}{\sqrt{\ell (\ell +1)}} \varepsilon_{abc} {\hat x}_c \,, 
\quad {\hat x}_a {\hat x}_a = R \,,
\ee
and generate the full matrix algebra $\mbox{Mat}(2 \ell +1)$.  There are three natural derivations of functions, defined by the adjoint action of $su(2)$ on $S_F^2$:
\be
f \rightarrow ad X_a^{(2 \ell + 1)} f := \lbrack X_a^{(2 \ell + 1)} \,, f \rbrack \,, \quad f \in \mbox{Mat}(2 \ell +1) \,.  
\ee
In the limit $\ell\rightarrow\infty$, the functions $\hat{x}_a$ are identified with the standard coordinates $x_a$ on $\mathbb{R}^3$, restricted to the unit sphere, and the infinite-dimensional algebra ${\cal C}^\infty(S^2)$ of functions on the sphere is recovered. Also in this limit, the derivations $[X_a^{(2 \ell + 1)},\cdot]$ become the vector 
fields $-i{\cal L}_a = \varepsilon_{abc} x_a \partial_b$, induced by the usual action of $SO(3)$.

In similar manner the product space $S_F^2 \times S_F^2$ is defined to be the algebra of $\left ((2\ell_L +1)(2\ell_R +1) \right )$ matrices $\mbox{Mat}(2 \ell_L +1) (2 \ell_R + 1)$.  There are now six Hermitian ``coordinate functions''
\be
{\hat x}_a^L := \frac{i}{\sqrt{\ell_L (\ell_L +1)}} X_a^{(2\ell_L+1)} \otimes 1_{2 \ell_R +1} \,, \quad {\hat x}_a^R :=1_{2 \ell_L +1} \otimes \frac{i}{\sqrt{\ell_R (\ell_R +1)}} X_a^{(2\ell_R+1)}
\,, \quad a=1,2,3 \,.
\ee
which satisfy
\be
\lbrack {\hat x}_a^L \,, {\hat x}_b^L \rbrack = \frac{i}{\sqrt{\ell_L (\ell_L +1)}} \varepsilon_{abc} {\hat x}_c^L \,, 
\quad 
\lbrack {\hat x}_a^R \,, {\hat x}_b^R \rbrack = \frac{i}{\sqrt{\ell_R (\ell_R +1)}} \varepsilon_{abc} {\hat x}_c^R 
\,, \quad \lbrack {\hat x}_a^L \,, {\hat x}_b^R \rbrack = 0 \,.
\ee
\be
{\hat x}_a^L {\hat x}_a^L = 1 \,, \quad \quad {\hat x}_a^R {\hat x}_a^R = 1 \,.
\ee
and generate the full matrix algebra $\mbox{Mat}(2\ell_L +1)(2\ell_R +1)$. 

There are six natural derivations of functions, defined by the adjoint action of $su(2) \oplus su(2) = so(4)$ on $S_F^2 \times S_F^2$:
\be
f \rightarrow ad X_a^L f := \lbrack X_a^L \,, f \rbrack \,, \quad f \rightarrow ad X_a^R f := \lbrack X_a^R \,, f \rbrack \,, \quad 
f \in \mbox{Mat}(2 \ell_L +1)(2 \ell_R +1) \,.  
\ee
In the limit $\ell_L \,, \ell_R \rightarrow\infty$, $\hat{x}_a^L$ $\hat{x}_a^R$ and  are identified with the standard coordinates $x_a^L$ and $x_a^R$ on $\mathbb{R}^6$, restricted to  $S^2 \times S^2$, and the infinite-dimensional algebra ${\cal C}^\infty(S^2 \times S^2)$ of functions on  $S^2 \times S^2$ is recovered. Also in this limit, the derivations become the vector fields $-i{\cal L}_a^L = \varepsilon_{abc} x_a^L \partial_b^L$, $-i{\cal L}_a^R = \varepsilon_{abc} x_a^R \partial_b^R$ induced by the usual action of $SO(3) \times SO(3)$.

\subsection{$U(2)$ Gauge Theory ${\cal M} \times S_F^2$}

\setcounter{equation}{0}

\vskip 0.5 em

{\it  i. Gauge theory on ${\cal M} \times S_F^2$:}

\vskip 0.5 em

The relevant $SU(\cal{N})$ Yang-Mills theory has the action
\be
S = \int_{{\cal M} } \, \mbox{Tr}_{{\cal N}} \Big (
\frac{1}{4g^2} F_{\mu \nu}^\dagger F_{\mu \nu} + 
(D_\mu \phi_a)^\dagger (D_\mu \phi_a) \Big ) +
 \frac{1}{\tilde{g}^2} \mbox{Tr}_{{\cal N}} \big ( F_{ab}^\dagger F_{ab} \big )
 +  a^2  \mbox{Tr}_{{\cal N}} \big ( (\phi_a \phi_a + {\tilde b})^2 \big ) \,.
\label{eq:actionfirst}
\ee
Here, $\phi_a \,(a=1,2,3)$ are anti-Hermitian scalars, transforming in the adjoint of ${\rm SU(}{\cal N}{\rm )}$ and in the vector representation of an additional global $SO(3)$ symmetry, $D_\mu \phi_a  = \partial_\mu \phi_a + \lbrack A_\mu \,, \phi_a \rbrack$ are the covariant derivatives and $A_\mu$ are the $su({\cal N})$ valued anti-Hermitian gauge fields associated to the curvature $F_{\mu \nu}$. $F_{ab}$ is given as 
\be
F_{ab} := \lbrack \phi_a \,, \phi_b \rbrack - \varepsilon_{abc} \phi_c \,, 
\label{eq:curvaturefuzzy}
\ee
In above $a$, $\tilde{b}$, $g$ and $\tilde{g}$ are constants and $\mbox{Tr}_{{\cal N}} = {\cal N}^{-1} \mbox{Tr}$ denotes a normalized trace. 

This theory spontaneously develops extra dimensions in the form of fuzzy spheres \cite{Aschieri:2006uw}. The potential terms for the scalars are 
positive definite, and the solutions 
\be
F_{ab} = 0 \,, \quad - \phi_a \phi_a = {\tilde b}
\label{eq:minimum1}
\ee
are evidently a global minima. Most general solution to this equation is not known. However depending on the values taken by the parameter ${\tilde b}$, a large class of solutions has been found in \cite{Aschieri:2006uw}. Here we restrict ourselves to the simplest situation.Taking the value of ${\tilde b}$ as the quadratic Casimir of an irreducible representation of ${\rm SU(2)}$ labeled by $\ell$, ${\tilde b} = \ell (\ell + 1)$ with $2\ell\in\mathbb{Z}$ and assuming further that the dimension ${\cal N}$ of the matrices $\phi_a$ is $(2 \ell +1) n$, (\ref{eq:minimum1}) is solved by the configurations of the form 
\be
\phi_a = X_a^{(2 \ell + 1)} \otimes {\bf 1}_n \,,
\label{eq:minimumsol}
\ee
where $X_a^{(2 \ell + 1)}$ are the (anti-Hermitian) generators of ${\rm SU(2)}$ in the irreducible representation $\ell$, which has dimension $2\ell+1$. We observe that this vacuum configuration spontaneously breaks the ${\rm U}({\cal N})$ down to ${\rm U}(n)$ which is the commutant of $\phi_a$ in (\ref{eq:minimumsol}).

Fluctuations about the vacuum (\ref{eq:minimumsol}) may be written as
\be
\phi_a = X_a + A_a \,, 
\label{eq:config1} 
\ee
where $A_a \in u(2\ell+1)\otimes u(n)$ and we have used the short-hand notation $X_a^{(2 \ell + 1)} \otimes {\bf 1}_n =: X_a$.  Then $A_a$ $(a=1,2,3)$ may be interpreted as three components of a ${\rm U}(n)$ gauge field on the fuzzy sphere $S_F^2$. $\phi_a$ are indeed the ``covariant coordinates'' on $S_F^2$ and $F_{ab}$ is the field strength, which takes the form
\be
F_{ab} = \lbrack X_a \,, A_b \rbrack - \lbrack X_b \,, A_a \rbrack + \lbrack A_a \,, A_b \rbrack - \varepsilon_{abc} A_c \,.
\ee
when expressed in terms of the gauge fields $A_a$.

To summarize, with (\ref{eq:config1}) the action in (\ref{eq:actionfirst}) takes the form of a ${\rm U(n)}$ gauge theory on ${\cal M} \times S_F^2(2 \ell + 1)$ with the gauge field components $A_M({\hat y}) = (A_\mu({\hat y}) \,, A_a({\hat y}))\in  u(n)\otimes u(2\ell+1)$ and field strength tensor (${\hat y}$ are a set of coordinates for the noncommutative manifold ${\cal M}$)
\begin{eqnarray}
F_{\mu\nu} &=& \partial_\mu A_\nu - \partial_\nu A_\mu + [A_\mu,A_\nu]  \nn \\
F_{\mu a} &=& D_\mu \phi_a = \partial_\mu \phi_a + [A_\mu, \phi_a ] \\
F_{ab} &=& [\phi_a,\phi_b] - \epsilon_{abc}\phi_c \nn \,.
\end{eqnarray}

{\it ii. The ${\rm SU(2)}$-Equivariant Gauge Field :}

\vskip 1em

Let us focus on the case of a ${\rm U(2)}$ gauge theory on ${\cal M}\times S_F^2$. The construction of the most general ${\rm SU(2)}$-equivariant gauge field on  $S_F^2$ can be performed as follows \cite{Seckin-Derek}: 

We pick the symmetry generators $\omega_a$ which generate $SU(2)$ rotations upto $U(2)$ gauge transformations. Accordingly, we choose
\be
\omega_a = X_a^{(2\ell+1)} \otimes {\bf 1}_2 - {\bf 1}_{2 \ell +1} \otimes \frac{i\sigma^a}{2} \,, \quad \omega_a\in u(2)\otimes u(2\ell+1) \,, \mbox{for} \,  a=1,2,3
\ee
These $\omega_a$ are the generators of the representation $\underline{1/2} \otimes \underline{\ell}$ of ${\rm SU(2)}$, 
where by $\underline{m}$ we denote the spin $m$ representation of ${\rm SU(2)}$ of dimension $2m+1$.  
${\rm SU(2)}$-equivariance of the theory requires the fulfillment of the symmetry constraints,
\be
\lbrack \omega_a \,, A_\mu \rbrack = 0  \,, \quad \lbrack \omega_a, \phi_b \rbrack = \epsilon_{abc} \phi_c,
\label{eq:vector}
\ee
on the gauge field and a consistency condition on these constraints is $\lbrack \omega_a, \omega_b \rbrack = \varepsilon_{abc} \omega_c$ which 
is readily satisfied by our choice of $\omega_a$.

The solutions to these constraints are obtained using the representation theory of $SU(2)$. The adjoint action of $\omega$ expands into the Clebsch-Gordan
series, whose relevant part reads
\be
(\underline{1/2} \otimes \underline{\ell}) \otimes (\underline{1/2} \otimes \underline{\ell}) = 2\,\underline{0} \oplus 4\,\underline{1} \oplus\dots \,.
\ee
Thus, the set of solutions to equations in (\ref{eq:vector}) are two and four-dimensional respectively. The fields are conveniently parametrized as
\be
A_\mu = \frac{1}{2}Q a_\mu({\hat y}) + \frac{1}{2}i b_\mu({\hat y}) \,,
\label{eq:amu}
\ee
\be
A_a = \frac{1}{2}\varphi_1({\hat y})[X_a,Q] + \frac{1}{2} (\varphi_2({\hat y})-1) Q[X_a,Q] 
+ i \frac{1}{2} \varphi_3({\hat y}) \frac{1}{2} \{ \hat{X}_a, Q \} +
\frac{1}{2} \varphi_4({\hat y}) \hat{\omega}_a,,
\label{eq:eqvansatz}
\ee
with $\phi_a = X_a + A_a$ and $a_\mu$, $b_\mu$ are Hermitian ${\rm U(1)}$ gauge fields, $\varphi_i$ are Hermitian scalar fields over ${\cal M}$, the curly brackets denote anti-commutators throughout, and 
\be
\hat{X}_a := \frac{1}{\ell+1/2} X_a \,, \quad {\hat \omega}_a := \frac{1}{\ell+1/2} \omega_a.
\ee
They contain, in addition to the  $\mbox{Mat}{2(2 \ell +1)} $ identity matrix, the only non-trivial rotational invariant under $\omega$, which is  
\begin{equation}
Q := \frac{X_a\otimes\sigma^a - i/2}{\ell+1/2} \,, \quad Q^\dagger = - Q \,,
\quad Q^2 = - {\bf 1}_{2(2 \ell +1)} \,.
\end{equation}
Indeed, $Q$ is the fuzzy version of $q := i {\bf \sigma} \cdot {\bf x}$ and converges to it in the $\ell \rightarrow \infty$ limit.

\subsection{Explicit Formulae}

\setcounter{equation}{0}

In this appendix, we list the explicit expressions for $P_1^{L \pm}$, $P_2^L$ and $P_3^L$, $ {\tilde P}_2^L$, $ {\tilde P}_3^L$,
$P_1^{R \pm}$, $P_2^R$ and $P_3^R$, $ {\tilde P}_2^R$, $ {\tilde P}_3^R$, $T_1^{L,R}, T_2^{L,R}, {\tilde T}_2^{L,R}, T_3^{L,R},$, $R_1^L$, $R_2^L$ and ${\tilde R}_1^L$, ${\tilde R}_2^L$  and $R_1^R$, $R_2^R$ and ${\tilde R}_1^R$, ${\tilde R}_2^R$, 
which were introduced for brevity of notation in section 3. 

We have
\beqa
P_1^L &=& \frac{\ell_L^2+\ell_L-1/4}{(\ell_L+1/2)^2}\chi_3 + \frac{1}{\ell_L+1/2}\chi_4 \,, \\
P_1^{L \prime} &=& \frac{\ell_L^2+\ell_L-1/4}{(\ell_L+1/2)^2}\chi_3^\prime + \frac{1}{\ell_L+1/2}\chi_4^\prime 
\eeqa
\be
P_2^L = (1-\chi_3)\left( 1 + \frac{\chi_4}{\ell_L+1/2} - \frac{\chi_3}{2(\ell_L+1/2)^2} \right) 
 - \chi_3^\prime \left (\frac{\chi_4^\prime}{\ell_L + 1/2} - \frac{\chi_3^\prime}{2(\ell_L + 1/2)^2}
\right) \,,
\ee
\beqa
P_3^L = \frac{\ell_L (\ell_L + 1)}{(\ell_L+1/2)^2} \left( \chi_3^2-2\chi_3
\right) + \chi_4^2 + 2\frac{\ell_L^2+\ell_L-1/4}{\ell_L+1/2} \chi_4 
+ \frac{\ell_L (\ell_L +1)}{(\ell_L +1/2)^2} \chi_3^{\prime 2} + \chi_4^{\prime 2} 
\,.
\eeqa
\be
P_1^{L \pm} = P_1^L \pm P_1^{L \prime} \,.
\ee
\beqa
P_1^R &=& \frac{\ell_R^2+\ell_L-1/4}{(\ell_R+1/2)^2}\lambda_3 + \frac{1}{\ell_R+1/2}\lambda_4 \,, \\
P_1^{R \prime} &=& \frac{\ell_R^2+\ell_R-1/4}{(\ell_R+1/2)^2}\lambda_3^\prime + \frac{1}{\ell_R+1/2}\lambda_4^\prime \,,
\eeqa
\be
P_2^R = (1-\lambda_3)\left( 1 + \frac{\lambda_4}{\ell_R+1/2} - \frac{\lambda_3}{2(\ell_R+1/2)^2} \right) - \lambda_3^\prime \left (\frac{\lambda_4^\prime}{\ell_R + 1/2} - \frac{\lambda_3^\prime}{2(\ell_R + 1/2)^2} \right ) \,,
\ee
\be
P_3^R = \frac{\ell_R (\ell_R +1)}{(\ell_R+1/2)^2} \left( \lambda_3^2-2\lambda_3
\right) + \lambda_4^2 + 2\frac{\ell_R^2+\ell_R-1/4}{\ell_R+1/2} \lambda_4 +
\frac{\ell_R (\ell_R +1)}{(\ell_R +1/2)^2} \lambda_3^{\prime 2} + \lambda_4^{\prime 2} \,,
\ee
\be
P_1^{R \pm} = P_1^R \pm P_1^{R \prime} \,.
\ee
\begin{multline}
{\tilde P}_2^L = -\left (1 + \frac{1}{2 (\ell_L + 1/2)^2} \right ) \chi_3^\prime + \frac{1}{(\ell_L + 1/2)} \chi_4^\prime + \frac{1}{(\ell_L + 1/2)^2} \chi_3 \chi_3^\prime  \\
- \frac{1}{(\ell_L + 1/2)}  \left (\chi_3 \chi_4^\prime + \chi_3^\prime \chi_4 \right) \,,
\end{multline}
\be
{\tilde P}_3^L = \frac{2 (\ell_L^2 +\ell_L -\frac{1}{4})}{(\ell_L + 1/2)} \chi_4^\prime + \frac{2\ell_L(\ell_L+1)}{(\ell_L + 1/2)^2} (\chi_3 - 1)\chi_3^\prime + 2\chi_4 \chi_4^\prime \,,
\ee

\begin{multline}
{\tilde P}_2^R = -\left (1 + \frac{1}{2 (\ell_R + 1/2)^2} \right ) \lambda_3^\prime + \frac{1}{(\ell_R + 1/2)} \lambda_4^\prime + \frac{1}{(\ell_R + 1/2)^2} \lambda_3 \lambda_3^\prime  \\
- \frac{1}{(\ell_R + 1/2)}  \left (\lambda_3\lambda_4^\prime + \lambda_3^\prime \lambda_4 \right) \,,
\end{multline}
\be
{\tilde P}_3^R = \frac{2 (\ell_R^2 +\ell_R -\frac{1}{4})}{(\ell_R + 1/2)} \lambda_4^\prime + \frac{2\ell_R (\ell_R+1)}{(\ell_R + 1/2)^2} (\lambda_3 - 1) \lambda_3^\prime 
+ 2 \lambda_4 \lambda_4^\prime \,,
\ee

\be
T_1^L = 4 \frac{\ell_L (\ell_L+1)(\ell_L^2+\ell_L - 1/4)}{(\ell_L+1/2)^4} \,, 
\ee

\begin{multline}
T_2^L =  2\frac{\ell_L (\ell_L +1)}{(\ell_L+1/2)^2} \left ( (P_1^{L +})^2 - \frac{\ell_L^2+\ell_L-1/4}{(\ell_L+1/2)^2} (P_2^L + {\tilde P}_2^L)
+ \frac{1}{2(\ell_L+1/2)^2} (P_3^L + {\tilde P}_3^L) \right ) \\
+ \frac{1}{(\ell_R + 1/2)} \frac{1}{(\ell_L + 1/2)^2} \Bigg ( \ell_L (\ell_L +1) (P_1^{L +})^2 + \frac{2 \ell_L (\ell_L +1)(\ell_L^2+\ell_L-1/4)}{(\ell_L+1/2)^2}
\left ( 1 - \frac{1}{2} (P_2^L + {\tilde P}_2^L) \right ) \\
+ \frac{1}{2 (\ell_L+1/2)} (P_3^L + {\tilde P}_3^L) \Bigg) \,.
\end{multline}

\begin{multline}
{\tilde T}_2^L =  2\frac{\ell_L (\ell_L +1)}{(\ell_L+1/2)^2} \left ( (P_1^{L -})^2 - \frac{\ell_L^2+\ell_L-1/4}{(\ell_L+1/2)^2} (P_2^L - {\tilde P}_2^L)
+ \frac{1}{2(\ell_L+1/2)^2} (P_3^L - {\tilde P}_3^L) \right ) \\
+ \frac{1}{(\ell_R + 1/2)} \frac{1}{(\ell_L + 1/2)^2} \Bigg ( - \ell_L (\ell_L +1) (P_1^{L -})^2 - \frac{2 \ell_L (\ell_L +1)(\ell_L^2+\ell_L-1/4)}{(\ell_L+1/2)^2}
\left ( 1 - \frac{1}{2} ({\tilde P}_2^L - P_2^L) \right ) \\
+ \frac{1}{2 (\ell_L+1/2)} ({\tilde P}_3^L - P_3^L ) \Bigg) \,.
\end{multline}

\begin{multline}
T_3^L = \frac{1}{2 (\ell_L+1/2)^4} \Bigg ( \ell_L (\ell_L +1) (\ell_L^2+\ell_L-1/4) \left ( (P_2^L)^2 + ({\tilde P}_2^L)^2 \right ) +
\frac{1}{4} (\ell_L^2+\ell_L+3/4) \left ( (P_3^L)^2 + ({\tilde P}_3^L)^2 \right ) \\ 
- \ell_L(\ell_L + 1) ( P_2^L P_3^L +  {\tilde P}_2^L {\tilde P}_3^L
) \Bigg ) + \frac{1}{2} \frac{1}{(\ell_R + 1/2)} \frac{1}{(\ell_L+1/2)^3}  \Bigg (  \frac{\ell_L (\ell_L +1)(\ell_L^2+\ell_L-1/4)}{(\ell_L+1/2)} 
P_2^L {\tilde P}_2^L \\ 
+ \frac{1}{4} \frac{(\ell_L^2+\ell_L+3/4) }{(\ell_L+1/2)} P_3^L {\tilde P}_3^L - \frac{1}{2} ( P_2^L {\tilde P}_3^L 
+ {\tilde P}_2^L P_3^L ) \Bigg ) \,.
\end{multline}

\be
T_1^R = 4 \frac{\ell_R(\ell_R+1)(\ell_L^R+\ell_R - 1/4)}{(\ell_R+1/2)^4} \,, 
\ee

\begin{multline}
T_2^R =  2\frac{\ell_R (\ell_R +1)}{(\ell_R+1/2)^2} \left ( (P_1^{R +})^2 - \frac{\ell_R^2+\ell_R-1/4}{(\ell_R+1/2)^2} (P_2^R + {\tilde P}_2^R)
+ \frac{1}{2(\ell_R+1/2)^2} (P_3^R + {\tilde P}_3^R) \right ) \\
+ \frac{1}{(\ell_L + 1/2)} \frac{1}{(\ell_R + 1/2)^2} \Bigg ( \ell_R (\ell_R +1) (P_1^{R +})^2 + \frac{2 \ell_R (\ell_R +1)(\ell_R^2+\ell_R-1/4)}{(\ell_R+1/2)^2}
\left ( 1 - \frac{1}{2} (P_2^R + {\tilde P}_2^R) \right ) \\
+ \frac{1}{2 (\ell_R+1/2)} (P_3^R + {\tilde P}_3^R) \Bigg) \,.
\end{multline}

\begin{multline}
{\tilde T}_2^R =  2\frac{\ell_R (\ell_R +1)}{(\ell_R+1/2)^2} \left ( (P_1^{R -})^2 - \frac{\ell_R^2+\ell_R-1/4}{(\ell_R+1/2)^2} (P_2^R - {\tilde P}_2^R)
+ \frac{1}{2(\ell_R+1/2)^2} (P_3^R - {\tilde P}_3^R) \right ) \\
+ \frac{1}{(\ell_L + 1/2)} \frac{1}{(\ell_R + 1/2)^2} \Bigg ( - \ell_R (\ell_R +1) (P_1^{R -})^2 - \frac{2 \ell_R (\ell_R +1)(\ell_R^2+\ell_R-1/4)}{(\ell_R+1/2)^2}
\left ( 1 - \frac{1}{2} ({\tilde P}_2^R - P_2^R) \right ) \\
+ \frac{1}{2 (\ell_R+1/2)} ({\tilde P}_3^R - P_3^R ) \Bigg) \,.
\end{multline}

\begin{multline}
T_3^R = \frac{1}{2 (\ell_R+1/2)^4} \Bigg ( \ell_R (\ell_R +1) (\ell_R^2+\ell_R-1/4) \left ( (P_2^R)^2 + ({\tilde P}_2^R)^2 \right ) +
\frac{1}{4} (\ell_R^2+\ell_R+3/4) \left ( (P_3^R)^2 + ({\tilde P}_3^R)^2 \right ) \\ 
- \ell_R(\ell_R + 1) ( P_2^R P_3^R +  {\tilde P}_2^R {\tilde P}_3^R
) \Bigg ) + \frac{1}{2} \frac{1}{(\ell_L + 1/2)} \frac{1}{(\ell_R +1/2)^3}  \Bigg (  \frac{\ell_R (\ell_R +1)(\ell_R^2+\ell_R-1/4)}{(\ell_R+1/2)} 
P_2^R {\tilde P}_2^R \\ 
+ \frac{1}{4} \frac{(\ell_R^2+\ell_R+3/4) }{(\ell_R+1/2)} P_3^R {\tilde P}_3^R - \frac{1}{2} ( P_2^R {\tilde P}_3^R 
+ {\tilde P}_2^R P_3^R ) \Bigg ) \,.
\end{multline}

\begin{multline}
R_1^L = - \frac{1}{2} \left (2 ( \chi_1^2 + \chi_2^2)+ 2 (\chi_1^{\prime 2} + \chi_2^{\prime 2} ) - 1 \right ) - 
\frac{1}{4(\elll + \frac{1}{2})^2}\chi_3 - \left((\elll + \frac{1}{2})-\frac{1}{2(\elll + \frac{1}{2})}\right) \chi_4 \\ 
- \frac{4 \elll (\elll+1) - 2}{16 (\elll + \frac{1}{2})^2} (\chi_3^2 + \chi_3^{\prime 2}) - \frac{1}{4}(\chi_4^2
+ \chi_4^{\prime 2}) - \frac{1}{4(\elll + \frac{1}{2})}(\chi_3 \chi_4 + \chi_3^\prime \chi_4^\prime) \,,
\end{multline}
\begin{multline}
R_2^L = \frac{1}{4(\elll + \frac{1}{2})} \left (2 ( \chi_1^2 + \chi_2^2)+ 2 (\chi_1^{\prime 2} + \chi_2^{\prime 2} ) - 1 \right )  -
\left( (\elll + \frac{1}{2}) - \frac{3}{4(\elll + \frac{1}{2})}\right) \chi_3 - \frac{1}{2} \chi_4 \\
- \frac{1}{16(\elll + \frac{1}{2})^3}(\chi_3^2 + \chi_3^{\prime 2}) 
- \left( \frac{1}{2} - \frac{1}{4(\elll + \frac{1}{2})^2} \right)(\chi_3 \chi_4 + \chi_3^\prime \chi_4^\prime) 
- \frac{1}{4(\elll + \frac{1}{2})}(\chi_4^2 + \chi_4^{\prime 2}) \,.
\end{multline}

\begin{multline}
{\tilde R}_1^L = - \frac{1}{2} \left (2 ( \chi_1^2 + \chi_2^2) - 2 (\chi_1^{\prime 2} + \chi_2^{\prime 2} ) \right ) - 
\frac{1}{4(\elll + \frac{1}{2})^2}\chi_3^\prime - \left((\elll + \frac{1}{2})-\frac{1}{2(\elll + \frac{1}{2})}\right) \chi_4^\prime \\ 
- \frac{4 \elll (\elll+1) - 2}{16 (\elll + \frac{1}{2})^2} ( 2 \chi_3 \chi_3^\prime) - \frac{1}{4}(2 \chi_4 \chi_4^\prime)
- \frac{1}{4(\elll + \frac{1}{2})}(\chi_3 \chi_4^\prime + \chi_3^\prime \chi_4) \,,
\end{multline}
\begin{multline}
{\tilde R}_2^L = \frac{1}{4(\elll + \frac{1}{2})} \left (2 ( \chi_1^2 + \chi_2^2) - 2 (\chi_1^{\prime 2} + \chi_2^{\prime 2} ) \right ) 
- \left( (\elll + \frac{1}{2}) - \frac{3}{4(\elll + \frac{1}{2})}\right) \chi_3^\prime - \frac{1}{2} \chi_4^\prime \\
- \frac{1}{16(\elll + \frac{1}{2})^3}(2 \chi_3 \chi_3^\prime) 
- \left( \frac{1}{2} - \frac{1}{4(\elll + \frac{1}{2})^2} \right)(\chi_3 \chi_4^\prime + \chi_3^\prime \chi_4) 
- \frac{1}{4(\elll + \frac{1}{2})}(2 \chi_4 \chi_4^\prime) \,.
\end{multline}

\begin{multline}
R_1^R = - \frac{1}{2} \left (2 ( \lambda_1^2 + \lambda_2^2)+ 2 (\lambda_1^{\prime 2} + \lambda_2^{\prime 2} ) - 1 \right ) - 
\frac{1}{4(\ellr + \frac{1}{2})^2}\lambda_3 - \left((\ellr + \frac{1}{2})-\frac{1}{2(\ellr + \frac{1}{2})}\right) \lambda_4 \\ 
- \frac{4 \ellr (\ellr+1) - 2}{16 (\ellr+ \frac{1}{2})^2} (\lambda_3^2 + \lambda_3^{\prime 2}) - \frac{1}{4}(\lambda_4^2
+ \lambda_4^{\prime 2}) - \frac{1}{4(\ellr + \frac{1}{2})}(\lambda_3 \lambda_4 + \lambda_3^\prime \lambda_4^\prime) \,,
\end{multline}
\begin{multline}
R_2^R = \frac{1}{4(\ellr + \frac{1}{2})} \left (2 ( \lambda_1^2 + \lambda_2^2)+ 2 (\lambda_1^{\prime 2} + \lambda_2^{\prime 2} ) - 1 \right )  -
\left( (\ellr + \frac{1}{2}) - \frac{3}{4(\ellr + \frac{1}{2})}\right) \lambda_3 - \frac{1}{2} \lambda_4 \\
- \frac{1}{16(\ellr + \frac{1}{2})^3}(\lambda_3^2 + \lambda_3^{\prime 2}) 
- \left( \frac{1}{2} - \frac{1}{4(\ellr + \frac{1}{2})^2} \right)(\lambda_3 \lambda_4 + \lambda_3^\prime \lambda_4^\prime) 
- \frac{1}{4(\ellr + \frac{1}{2})}(\lambda_4^2 + \lambda_4^{\prime 2}) \,.
\end{multline}

\begin{multline}
{\tilde R}_1^R = - \frac{1}{2} \left (2 ( \lambda_1^2 + \lambda_2^2) - 2 (\lambda_1^{\prime 2} + \lambda_2^{\prime 2} ) \right ) - 
\frac{1}{4(\ellr + \frac{1}{2})^2}\lambda_3^\prime - \left((\ellr + \frac{1}{2})-\frac{1}{2(\ellr + \frac{1}{2})}\right) \lambda_4^\prime \\ 
- \frac{4 \ellr (\ellr + 1) - 2}{16 (\ellr + \frac{1}{2})^2} ( 2 \lambda_3 \lambda_3^\prime) - \frac{1}{4}(2 \lambda_4 \lambda_4^\prime)
- \frac{1}{4(\ellr + \frac{1}{2})}(\lambda_3 \lambda_4^\prime + \lambda_3^\prime \lambda_4) \,,
\end{multline}
\begin{multline}
{\tilde R}_2^R = \frac{1}{4(\ellr + \frac{1}{2})} \left (2 ( \lambda_1^2 + \lambda_2^2) - 2 (\lambda_1^{\prime 2} + \lambda_2^{\prime 2} ) \right ) 
- \left( (\ellr + \frac{1}{2}) - \frac{3}{4(\ellr + \frac{1}{2})}\right) \lambda_3^\prime - \frac{1}{2} \lambda_4^\prime \\
- \frac{1}{16(\ellr + \frac{1}{2})^3}(2 \lambda_3 \lambda_3^\prime) 
- \left( \frac{1}{2} - \frac{1}{4(\ellr + \frac{1}{2})^2} \right)(\lambda_3 \lambda_4^\prime + \lambda_3^\prime \lambda_4) 
- \frac{1}{4(\ellr + \frac{1}{2})}(2 \lambda_4 \lambda_4^\prime) \,,
\end{multline}

\be
S_1 = \frac{\ell_L (\ell_L +1) \ell_R (\ell_R +1) }{(\ell_L+1/2)^2 (\ell_R+1/2)^2} \,, 
\ee

\be
S_2^L = - \frac{1}{4} \frac{\ell_L (\ell_L +1) \ell_R (\ell_R +1) }{(\ell_L+1/2)^2 (\ell_R+1/2)^2} \left ( \frac{(\ell_R +\frac{3}{2})(\ell_R - \frac{1}{2})}{(\ell_R +\frac{1}{2})^2} +1 \right ) \,,
\ee

\be
{\tilde S}_2^L = - \frac{1}{2} \frac{\ell_L (\ell_L +1) (\ell_R^2 + \ell_R +\frac{3}{4}) }{(\ell_L+1/2)^2 (\ell_R+1/2)^2}
\ee

\be
S_3^L = - \frac{1}{2} \frac{\ell_L (\ell_L +1) \ell_R (\ell_R +1) }{(\ell_L+1/2)^2 (\ell_R+1/2)^3} =  - \frac{1}{2  (\ell_R+1/2) } S_1
\ee

\be
S_2^R = - \frac{1}{4} \frac{\ell_L (\ell_L +1) \ell_R (\ell_R +1) }{(\ell_L+1/2)^2 (\ell_R+1/2)^2} \left ( \frac{(\ell_L +\frac{3}{2})(\ell_L - \frac{1}{2})}{(\ell_L +\frac{1}{2})^2} +1 \right ) \,,
\ee

\be
{\tilde S}_2^R =  - \frac{1}{2} \frac{\ell_R (\ell_R +1) (\ell_L^2 + \ell_L +\frac{3}{4}) }{(\ell_L+1/2)^2 (\ell_R+1/2)^2}
\ee

\be
S_3^R = - \frac{1}{2} \frac{\ell_L (\ell_L +1) \ell_R (\ell_R +1) }{(\ell_L+1/2)^3 (\ell_R+1/2)^2} =  - \frac{1}{2  (\ell_L+1/2) } S_1
\ee

\end{document}